\documentstyle[12pt,epsf]{article}
\input{amssym.def}
\input{amssym.tex}
\catcode`\@=11
\@addtoreset{equation}{section}

\def\theequation{\thesection.\arabic{equation}}

\newtoks\@stequation

\def\subequations{\refstepcounter{equation}%
  \edef\@savedequation{\the\c@equation}%
  \@stequation=\expandafter{\theequation}
  \edef\@savedtheequation{\the\@stequation}
  \edef\oldtheequation{\theequation}%
  \setcounter{equation}{0}%
  \def\theequation{\oldtheequation\alph{equation}}}

\def\endsubequations{\setcounter{equation}{\@savedequation}%
  \@stequation=\expandafter{\@savedtheequation}%
  \edef\theequation{\the\@stequation}\global\@ignoretrue
  \vspace*{-12pt} \\}

\catcode`\@=11

\def\marginnote#1{}

\newcount\hour
\newcount\minute
\newtoks\amorpm
\hour=\time\divide\hour by60
\minute=\time{\multiply\hour by60 \global\advance\minute by-\hour}
\edef\standardtime{{\ifnum\hour<12 \global\amorpm={am}%
        \else\global\amorpm={pm}\advance\hour by-12 \fi
        \ifnum\hour=0 \hour=12 \fi
        \number\hour:\ifnum\minute<10 0\fi\number\minute\the\amorpm}}
\edef\militarytime{\number\hour:\ifnum\minute<10 0\fi\number\minute}
\def\draftlabel#1{{\@bsphack\if@filesw {\let\thepage\relax
   \xdef\@gtempa{\write\@auxout{\string
      \newlabel{#1}{{\@currentlabel}{\thepage}}}}}\@gtempa
   \if@nobreak \ifvmode\nobreak\fi\fi\fi\@esphack}
        \gdef\@eqnlabel{#1}}
\def\@eqnlabel{}
\def\@vacuum{}
\def\draftmarginnote#1{\marginpar{\raggedright\scriptsize\tt#1}}

\def\draft{\oddsidemargin -.2truein
        \def\@oddfoot{\sl preliminary draft \hfil
        \rm\thepage\hfil\sl\today\quad\militarytime}
        \let\@evenfoot\@oddfoot \overfullrule 3pt
        \let\label=\draftlabel
        \let\marginnote=\draftmarginnote
   \def\@eqnnum{(\theequation)\rlap{\kern\marginparsep\tt\@eqnlabel}%
\global\let\@eqnlabel\@vacuum}  }

\def\xxx#1           {{hep-th/#1}}

\def\npb#1(#2)#3     {Nucl. Phys. {\bf B#1} (#2) #3 }
\def\rep#1(#2)#3     {Phys. Rept.{\bf #1} (#2) #3 }
\def\plb#1(#2)#3     {Phys. Lett. {\bf #1B} (#2) #3 }
\def\prl#1(#2)#3     {Phys. Rev. Lett.{\bf #1} (#2) #3 }
\def\prd#1(#2)#3     {Phys. Rev. {\bf D#1} (#2) #3 }
\def\ap#1(#2)#3      {Ann. Phys. {\bf #1} (#2) #3 }
\def\rmp#1(#2)#3     {Rev. Mod. Phys. {\bf #1} (#2) #3 }
\def\cmp#1(#2)#3     {Comm. Math. Phys. {\bf #1} (#2) #3 }
\def\mpla#1(#2)#3    {Mod. Phys. Lett. {\bf A#1} (#2) #3 }
\def\ijmp#1(#2)#3    {Int. J. Mod. Phys. {\bf A#1} (#2) #3 }
\def\cqg#1(#2)#3     {Class. Quant. Grav. {\bf #1} (#2) #3 }
\def\am#1(#2)#3      {Adv. Math. {\bf #1} (#2) #3 }
\def\im#1(#2)#3      {Invent. Math. {\bf #1} (#2) #3 }
\def\jhep#1(#2)#3    {J. High Energy Phys. {\bf #1} (#2) #3 }

\def\be{\begin{equation}}
\def\ee{\end{equation}}
\def\bea{\begin{eqnarray}}
\def\eea{\end{eqnarray}}

\def\W8{\bf W_8}
\def\W9{\bf W_9}
\def\A8{\bf A_8}
\def\A9{\bf A_9}
\def\A{\bf A}

\def\l{\lambda }
\def\E8E8{E_8\times E_8}
\def\spin32{Spin(32)/Z_2}
\def\f{\frac}
\def\1/2{\frac{1}{2}}
\def\l{\label}
\def\nn{\nonumber}

\begin{document}

\begin{flushright}
{\normalsize \ {\tt hep-th/9909178}} \\
{\normalsize \ DAMTP--1999--130 } \\
\end{flushright}
\vskip 0.5cm

\begin{center}\LARGE
{\bf Map of Heterotic and Type IIB Moduli \\
in 8 Dimensions}
\end{center}
\vskip 1.0cm
\begin{center}
{\large  M.C. Daflon Barrozo.\footnote{E-mail  address:
{\tt M.C.D.Barrozo@damtp.cam.ac.uk}}}

\vskip 0.5 cm
{\it Department of Applied Mathematics and Theoretical Physics\\
Cambridge University, Cambridge, England}
\end{center}

\vskip 1.0cm

\begin{center}
September, 1999
\end{center}

\vskip 1.0cm

\begin{abstract}
Explicit relations among moduli of the Heterotic and Type IIB string
theories in 8 dimensions are obtained. We identify the BPS states
responsible for gauge enhancements in the type IIB theory and their dual
partners in the Heterotic theory compactified with and without Wilson
lines. The masses of BPS states in Type IIB string theory compactified
on the base space of a elliptically fibred $K3$ are computed
explicitly for the special cases in which the complex structure of the
fibre is constant, ie, for constant scalar fields backgrounds.  
\end{abstract}

\newpage

\tableofcontents

\baselineskip=18pt

\section{Introduction}

Over the past couple of years non-perturbative aspects of
compactifications of Type IIB theory have been studied first in terms
of F-theory \cite{vafa1} compactifications and later in an explicitly
stringy form by means of including non-perturbative
$(p,q)$ 7-brane configurations in the background. The former has been
particularly important in the analysis of the duality with the
Heterotic string.

Compactifications of F-theory on elliptic Calabi-Yau
two-folds ($K3$), three-folds and four-folds have been argued to be
dual to certain compactifications of the Heterotic string to 8, 6 and 4
dimensions. The simplest case to consider is the compactifications of F-theory to 8
dimensions on an elliptic $K3$. This theory is believed to be dual to the
Heterotic string on $T^2$. The $ADE$ pattern of gauge symmetry
enhancement in the Heterotic string \cite{witten1,ginsparg} is
reproduced in F-theory by the pattern of collapsible holomorphic
two spheres in $K3$ \cite{vafa2}. The location in the moduli space
where the symmetries occur are provided by the zeros of the discriminant of the
$K3$ surface. This establishes the duality at a geometrical level.

The other approach that has emerged \cite{sen1,johansen,
gab1,gab2,imamura,zwiebach1,zwiebach2,zwiebach3,dewolfe} is type IIB theory
compactified on a sphere in the presence of non-local
7-branes which extend in the uncompactified directions and appear as
singular points on the sphere. The presence of the $ADE$ series of algebras arising on
7-branes configurations were explored in
refs\cite{johansen,gab1,gab2,imamura,zwiebach1} where it was
shown how $(p,q)$-strings and string junctions stretched between the
7-branes correspond to vector bosons of the eight dimensional gauge
theory. Other algebras organised in terms of the conjugacy classes of
$SL(2,Z)$ have been identified in terms of the strings and junctions
connecting the 7-branes \cite{zwiebach2, zwiebach3, dewolfe}. Some
connections among elements in the Type IIB theory and their duals in the
Heterotic theory in 8 dimensions have appeared in the
literature\cite{imamura2}. However
all identifications have been at a qualitative level, in the sense that
the masses of the gauge bosons that are supposed to become
massless at specific points in the moduli space to generate the
relevant gauge group had not been checked explicitly. We will compute
explicitly the masses of the relevant gauge bosons in many examples in this paper.

The masses of BPS $(p,q)$-strings have been
computed explicitly in two cases. The first was the work of Sen
\cite{sen1} who used the results of Seiberg-Witten for $N=2$ supersymmetric $SU(2)$ gauge
theory with 4 quark flavours. He computed the masses of BPS states in the
limit where the 24 7-branes are grouped into four groups of 8 branes
yielding an $SO(8)^4$ symmetry. In \cite{sen2} Sen introduced a new
mass formula which showed how the masses above originated from the
mass formula of open strings stretched between 7-branes. Later,
Lerche and Stieberger \cite{lerche1} suggested a
generalisation of the formula used by Sen in terms of a contribution by the fundamental
period of the implicit F-theory $K3$. They argued that this factor
would not have any effect in the rigid limit considered by Sen but
would provide the correct normalisation in the general case. The
masses of the BPS gauge fields responsible for the
geometrical enhancement of $U(1)^2_G \rightarrow SU(3)_G$ were
computed and an exact match with the Heterotic string mass formula was
obtained. However, only one example and with no Wilson lines was considered.

In this paper we use the mass formula of ref\cite{lerche1} to compute
the mass of various gauge bosons potentially responsible for several
gauge enhancements in the type IIB side. This takes the description of
non perturbative Type IIB theory to a more quantitative level. We also
compare the masses of the BPS gauge bosons in Type IIB with their
Heterotic dual partners and find complete agreement. We analyse
examples with both zero and non zero Wilson lines. The explicit map of
the geometrical and Wilson lines moduli in the Heterotic theory to the
moduli describing the position of the 7-branes in the sphere is
obtained in several examples.

We focus on the branches of moduli space where the massless scalar fields in
Type IIB are constant. There are three such branches in the moduli
space \cite{sen1, mukhi}. We refer to them as Branches 0, I and
II. Branch 0 exist for any constant value of the scalar fields and allow
for only one symmetry group, namely, $SO(8)^4$. Branches I and II require
special values for the scalar fields and allow
for a more complex set of gauge symmetries. The
moduli space of these branches have dimensions 1, 5, and 8,
respectively. We will consider,
in particular, one-parameter families of curves living in their
multidimensional moduli spaces. Once we define a one-parameter family
in one of the branches of Type IIB by choosing a pattern of gauge enhancement we
compute the mass of the BPS gauge bosons responsible for the
symmetries. The next step is to determine the dual family in the
Heterotic theory that must
have not only the same pattern of gauge enhancements but also the masses
of the BPS gauge bosons must match everywhere in the moduli space
(See Fig 1). This procedure is carried out for a number of examples and
the explicit duality maps between the dual families are obtained.  

\begin{figure}[htb]
\epsfysize=5cm
\centerline{\epsffile{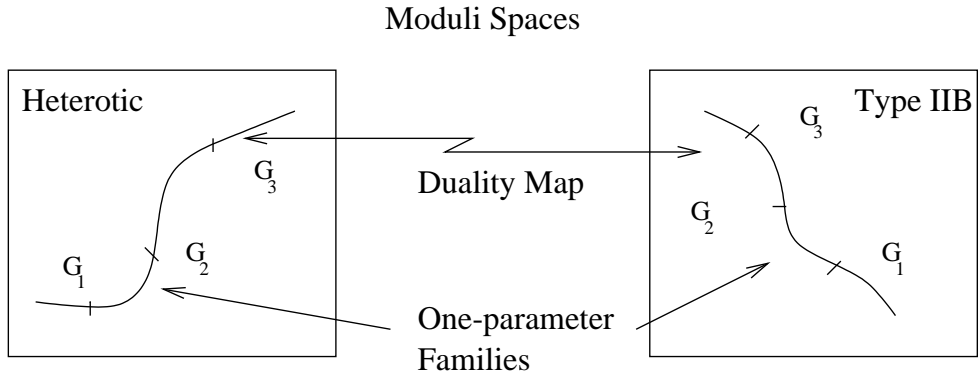}}
\caption{One-to-one duality map for one-parameter families in Type IIB
and Heterotic theories. The gauge group, $G_i$, enhancement pattern
must agree as well as the mass of the BPS states.}
\end{figure}

This paper is organised as follows. In section 2 we review
results for the Heterotic theory compactified on $T^2$. We also obtain
an expression for masses of BPS states in the presence of general
Wilson lines that generalises the standard holomorphic expression in
terms of the Kahler structure, $T$, and complex structure, $U$, of the
torus. This expression will  play a fundamental role in establishing
the duality map in Section 4. In section 3 we review the basics of type IIB
compactifications on the sphere in the presence of  7-branes in order
to establish our conventions. In section 4 we compute the masses of
a number of BPS states in the two branches of constant $\tau$ on the type IIB
side and compare with the masses of the Heterotic theory duals
obtained in Section 2. Explicit maps between the Heterotic and
Type IIB moduli are derived. In section 5 we  present our
conclusions. In Appendix A we show how to generalise some
hypergeometric relations used in the paper. And finally, in Appendix B
we obtain the explicit relation among certain moduli in the Heterotic
theory that are required elsewhere in the paper.

\section{Heterotic String on $T^2$}
Consider the Heterotic String compactified on a torus, $T^2$, down to 8
dimensions along directions $x^8$ and $x^9$. In this section it will not be necessary to specify which
Heterotic theory we are working with.

The expression for the left and right moving momenta with Wilson lines is
given by

\begin{equation}
(P_{Ri} | P_{Li}) = (p_i - (g_{ij} - B_{ij})\frac{w ^j}{\alpha '}  | p_i +
(g_{ij} + B_{ij})\frac{w ^j}{\alpha '};\; \sqrt{\frac{2}{\alpha
'}}{\bf q})
\end{equation}
where $m_i$ and $w^i$ are the KK momentum and winding numbers,
respectively, along $x^8$ and $x^9$. We define $p_i = m_i -\1/2 {\bf A}_i\cdot ({\bf A}_kw ^k) - {\bf A}_i\cdot
{\bf Q}$, ${\bf Q}$ is an element of either of the lattices $\Gamma
^{16}$ or $\Gamma ^8\bigoplus \Gamma ^8$ and ${\bf q} = {\bf Q} + w^k {\bf A_k}$. 

The mass spectrum is given by 
\be
M_h^2 = \f{P_L^2}{2} + \f{P_R^2}{2} +\f{2}{\alpha '}(N_L - 1) +\f{2}{\alpha '} (N_R - c_R) 
\ee
where $N_L$, $N_R$ are the left-, right-moving oscillator numbers, and
$c_R= 0 , \1/2$ depending on whether the right-moving fermions are
periodic $(R)$ or anti-periodic $(NS)$. We must impose the level
matching condition to obtain the physical states
\be
\f{P_L^2}{2} + \f{2}{\alpha '}(N_L - 1)=\f{P_R^2}{2} +\f{2}{\alpha '} (N_R - c_R),
\ee
BPS  states are given by the additional requirement that $N_R =
c_R$ \cite{harvey}. Therefore for states that are physical and BPS saturated we must
have
\be
P_L^2 - P_R^2 = \f{4}{\alpha '}(1- N_L),
\ee
or 
\be
\1/2 Q^2 + m_iw^i = 1-N_L.
\label{1}\ee
(summation convention). In this case we can write for the mass formula 
\bea
M_h^2 &=& P_{Ri}g^{ij}P_{Rj} \nn \\
&=& (p_i - (g_{il} - B_{il})\frac{w ^l}{\alpha '})g^{ij}(p_j - (g_{jk} - B_{jk})\frac{w ^k}{\alpha '})  .
\l{1a}
\eea

The massless states with $N_L = 1$ include the
8-dimensional metric, antisymmetric tensor, dilaton and gauge fields
which are the Cartan generators of the gauge group. We also have two scalars
in the adjoint of the gauge group which are the Wilson
lines. In the $N_L=0$ sector we have massless gauge fields associated
to the roots of the underlying gauge group. They must satisfy
\bea
P_R^2&=&0  \nn \\
P_L^2&=& \f{4}{\alpha '} \;\; \Rightarrow\;\; \1/2 Q^2 + m_iw^i = 1. 
\label{2}
\eea

The zero winding numbers sector gives the roots of the subgroup of
$E_8\times E_8$ or $SO(32)$ which is left unbroken by the Wilson
lines. Further gauge fields will appear in the non-zero winding numbers
sector for special values of the geometrical moduli of the torus, ie,
$g_{ij}$ and $B_{ij}$. We discuss these cases in more detail in later sections.

One of the goals of this paper is to use the duality between the
Heterotic strings on $T^2$ and Type IIB compactified on the base space of a
elliptic fibred $K3$. When comparing Heterotic string states with the dual states
in type IIB it turns out to be extremely convenient to rewrite the
Heterotic mass formula in a holomorphic or anti-holomorphic form. To do so we introduce the parameters 
\bea
U=U_1 + iU_2 &=& \f{g_{89}}{g_{99}} + i\f{\sqrt{g}}{g_{99}} \\ [2mm]
\label{3}
T^0=T^0_1+iT^0_2 &=&  \f{B}{2} + i\f{\sqrt{g}}{2}
\label{4}
\eea
where $g = g_{88}g_{99} - g_{89}^2$ and $B= B_{89}$. $T^0$ is the Kahler
structure and $U$ the complex structure of the torus. When dealing
with Wilson lines it becomes very convenient to introduce a modified version of the Kahler
structure. In the presence of Wilson lines we redefine \footnote{A
similar expression has been considered in compactifications of the
Heterotic string on $Z_N$ orbifolds \cite{mohaupt}. However, only for
the case of very specific Wilson lines.} 
\bea
U &=& U^0 \nonumber \label{4a} \\
T &=& (\f{B}{2} -\f{{\bf A_8}\cdot {\bf A_9}}{2} + \f{{\bf A_9}\cdot {\bf
A_9} g_{89}}{2g_{99}}) + i \f{\sqrt{g}}{2}(1 +
 \f{{\bf A_9}\cdot{\bf A_9} }{g_{99}}) \label{5} \\
Z &=& \1/2 (-\f{{\bf A_8}^2g_{99}^2 +{\bf A_9}^2(-g_{89} + g) +2{\bf
A_9}\cdot {\bf A_8} g_{89}g_{99}}{g_{99}^2})  - i \f{({\bf A_9}\cdot
{\bf A_9}g_{89}-{\bf A_8}\cdot {\bf A_9}g_{99})\sqrt{g}}{g_{99}^2}
\label{6} \\
n_i &=& m_i - {\bf A_i} \cdot {\bf Q} .
\label{7}
\eea
Note that $T \rightarrow T^0$ and $Z \rightarrow
0$ as we turn off the Wilson lines. The mass formula, eq(\ref{1a}),
becomes, in terms of the new variables,
\be
M_{BPS}=\frac{1}{2T_2^0U_2^0}|n_8 - U n_9 + T w^9 + (TU + Z)w^8|
\label{8}
\ee
Note that this expression for the mass of BPS states is valid for any
Wilson line. This mass formula turns out to be a very useful way of
rewriting the standard expression, eq(\ref{1a}), for compactifications
of the Heterotic string on $T^2$, particularly when there are Wilson
lines. It will play a major role in what follows.

\section{Type IIB on $S^2_s$}

Let us consider Type IIB theory compactified on a two-sphere, $S^2$, in
the presence of 24 parallel 7-branes which appear as points on
$S^2$. We refer to this punctured sphere as, $S^2_s$. The theory
possesses different strings labelled $(p,q)$ according to how
it is charged with respect to the $RR$ and $NSNS$
antisymmetric fields. The 7-branes of the theory are also labelled $(p,q)$,
according to what $(p,q)$-string can end on it.

In this convention the elementary string is $(1,0)$ and a D-string is
$(0,1)$. Each of the 24 7-branes has an associated  branch cut
depending on its type. We follow the conventions of \cite{gab2}. We
use three basic types of branes, ${\bf A}$, ${\bf B}$ and ${\bf
C}$, whose corresponding $(p,q)$ labels are $(1,0)$,
$(1,-1)$ and $(1,1)$ respectively. Across the corresponding cuts the
labels $(p,q)$ and $U$ change according to the monodromy matrices   
\bea
{\bf A}=& (1,0): \,\,K_A = &  T^{-1} = 
\pmatrix{1 & -1 \cr 0 & 1} \,, \nonumber\\ 
{\bf B}=& (1,-1): K_B = &   S T^{2} 
             = \pmatrix{0 & -1 \cr 1 & 2} \,, \\
{\bf C}=& (1,1): \,\, K_C = &   T^{2}S 
             = \pmatrix{2 & -1 \cr 1 & 0} \,,\nonumber
\eea  
where $S$ is the matrix  
\be
S = \pmatrix{ 0 & - 1 \cr 1 & 0 } \,.
\ee

All branes have their branch cut going upwards
vertically. ${\bf A}$-branes are represented by heavy dots, ${\bf B}$-branes by
empty boxes and ${\bf C}$-branes by empty circles (see Fig. 1). 
\begin{figure}[htb]
\epsfysize=5cm
\centerline{\epsffile{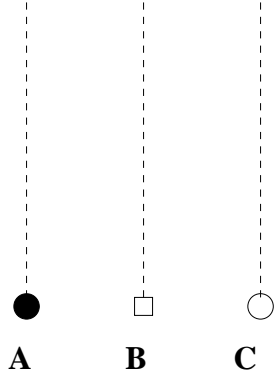}}
\caption{We start with a basic configuration where we have 16 ${\bf
A}$-branes, 4 ${\bf B}$-branes and 4 ${\bf C}$-branes. All branch cuts
are chosen to go upward.}
\end{figure}

A formula for the mass of $(p,q)$-strings in the background described
above has been derived by Sen \cite{sen1,sen2} and is given by

\be 
M_{IIB}(p,q)=\int_C
|\eta(\tau(z))^2\prod_{i=1}^{24}(z-z_i)^{-1/12}(p+q\tau(z))\, dz|\, .
\l{9}
\ee
For more general backgrounds there has been a suggestion
\cite{lerche1} that this mass formula should be generalised in order
to include additional F-theory data. Specifically, the authors of ref\cite{lerche1}
suggested that the mass formula should include the
fundamental period of $K3$, $\omega _0$, so that the mass formula becomes

\be
M_{BPS}^{IIB}= \frac{1}{|\omega _0|}\int_{z_i}^{z_j} |(p+q\tau(z)
)\eta (\tau (z) )^2\Delta (z)^{-\f{1}{12}}dz|,
\l{10}
\ee
where 
\be
\Delta = \prod _{i=1}^{24}(z-z_i).
\ee
Note that $\Delta= 4f^3 + 27g^2$ is the discriminant of the elliptic fibre, defined from 
the polynomial equation, $W(x,y,\xi)= y^2 - x^3 - f(\xi)x - g(\xi) =
0$, defining the elliptic $K3$ surface. The $z_i$'s are the
positions of the 24 branes on the sphere, $S^2_s$. 

The periods of a fibred $K3$ are given by

\be
\omega _i =\int_{\gamma_i}\frac{\!\!dx\, d\xi}{\partial_y W(x,y,\xi)}.
\l{11}
\ee  

\subsection{Branches of constant $\tau $}

From the expression for the j-function of the elliptic fibre
\be
j(\tau)= \f{4(24f)^3}{4f^3 + 27 g^2} ,
\ee
we see that there are three branches of constant $\tau$ \cite{sen1,mukhi}. We
have constant $\tau$ if $f^3 = \alpha g^2$, $f=0\; \& \; g\neq 0$
(Branch I) and $f\neq 0\; \&  \;  g=0$ (Branch II). It is
important that for each of these cases we can factorise the data in
the periods of $K3$ in a part depending on the elliptic fibre from that
depending on the base only. We summarise the results in Table 1.
\begin{table}[hth]
\begin{center}
\begin{tabular}{|l l|l|} \hline 
$f^3 =\alpha g^2$ & $\Rightarrow \Delta = (4\alpha +27)g^2
$ & 
$\omega _i = \int_{\gamma _i}\frac{dv}{(v^3+\alpha
^{1/2}v+1)^{1/2}}\int\frac{d\xi}{g^{1/6}}
=A(\tau)\int\frac{d\xi}{\Delta
^{1/12}}$   \\ [2mm] \hline
$f\neq 0 \; \& \; g=0$ & $\Rightarrow \Delta = 4f^3$ & 
$\omega _i = \int_{\gamma _i}\frac{dv}{(v(v^2+1))^{1/2}}\int\frac{d\xi}{f^{1/4}}
=A(i)\int\frac{d\xi}{\Delta
^{1/12}}$ \\ [2mm] \hline
$f = 0 \; \& \; g\neq 0$ & $\Rightarrow \Delta = 27g^2$ & 
$\omega _i = \int_{\gamma _i}\frac{dv}{(v^3+1)^{1/2}}\int\frac{d\xi}{g^{1/6}}
=A(e^{\frac{i\pi }{3}})\int\frac{d\xi}{\Delta
^{1/12}}$   \\ \hline 
\end{tabular}
\end{center}
\caption{Branches of constant $\tau$. $A(\tau)$ is a constant carrying information on the elliptic fibre.}
\end{table} 

For the three branches of constant $\tau$ we have an expression for the periods of
$K3$ in terms of the discriminant, $\Delta$, which is similar to
the one we have in the numerator of the the mass formula for
$(p,q)$-strings. Actually, the only subtle point in these expressions
in this case are the limits of
integration in the periods. In fact, we can write for constant $\tau$  
\be
M^{IIB}(p,q) =\biggl| \frac{(p+q\tau)\eta (\tau)^2}{A(\tau)}\biggr| \f{\int_{z_i}^{z_j}|\Delta
^{-\frac{1}{12}}|}{|\int_{\gamma_i} \Delta ^{-\frac{1}{12}}|}.
\ee

In the next section we compute the masses of BPS gauge fields and study
how they map under the Heterotic and Type IIB theories duality in the branches I and II. The
branch where $f^3 = \alpha g^2$ only has $SO(8)^4$ symmetry and we
will not consider it in this paper.

\section{The Duality Map}

In this section we will consider the masses of BPS states in Type IIB
theory in two branches of constant $\tau$, namely, Branch I and II. And then we identify the dual BPS
states in the Heterotic string and compare their masses. This will
allow us to identify explicitly the duality map for some moduli of the
two theories.

\subsection{Branch I - ${\bf \tau=e^{i\pi/3}}$}

In Branch I, $f=0 \;\; \& \;\; g\neq 0  \rightarrow \tau = e^{\frac{i\pi
}{3}}$, there are only 9 degrees of freedom. The 24 branes join up in
12 non local pairs of branes. The two branes forming each pair can only move together. We
will refer to them as a dynamical unit. Each dynamical unit is formed
by an ${\bf AC}$ pair. Their positions on $S^2_s$ are related though. In fact,
since the zeros of the discriminant indicate the position of the
branes on $S^2_s$ and $g(\xi)$ is a polynomial of degree 12, we have $\Delta =
27g^2 = 0$. This equation has 12 degrees of freedom. Moding out by
$SL(2,{\bf C})$ eliminates another 3 complex degrees of freedom giving
relations among the position of the branes on the sphere. 

These singularities can collide and yield a gauge enhancement
at the points where the discriminant of $K3$ vanishes. The pattern of gauge
enhancement in each of the branches of constant $\tau$ in Type IIB has
been studied in detail in ref\cite{gab1}. The basic gauge
enhancements that appear when $\tau= e^{\frac{i\pi}{3}}$ due to dynamical units colliding at the same point is given by
\be
U(1)^2 \rightarrow SU(3); \;\;\; U(1)^3 \rightarrow SO(8);  \;\;\;
U(1)^4 \rightarrow E_6; \;\;\; U(1)^5 \rightarrow E_8. 
\ee

In ref\cite{gab1} the authors qualitatively identified the gauge fields
responsible for the symmetries above. In ref\cite{zwiebach1} a
systematic procedure based in string junctions was developed giving
further evidence for the identification of these gauge fields. However
no explicit check of the mass for this gauge fields and its relation
to the heterotic duals has been obtained until recently. In
\cite{lerche1} (see also \cite{lerche2, lerche3}) the masses of
the gauge fields responsible for the enhancement of $E_8\times
E_8\times U(1)_G\times U(1)_G \rightarrow E_8\times E_8\times SU(3)_G$
were computed. They were shown to be identical to the mass of the heterotic duals.

The mass for BPS $(p,q)$-strings in this branch is given by
\be
M_{IIB}(p,q) = \biggl| \f{(p+\tau q)\eta (e^{i\pi /3})^2}{A(e^{i\pi /3})}\biggr| \f{
\int_{z_i}^{z_j} dz |\Delta (z)^{(-1/12)}|}{|\int_{\gamma_0} dz \Delta
(z)^{(-1/12)}|}.
\l{12}
\ee

We now compute the mass of BPS gauge fields in both the Heterotic and
Type IIB theories. We start by reviewing the results of
ref\cite{lerche1}. We will then turn on Wilson lines and analyse how
it effects the mass for the BPS gauge bosons.

\subsubsection*{a) The case with no Wilson lines: ${\bf E_8\times E_8 \times U(1)_G^2 \rightarrow E_8\times E_8
\times SU(3)_G}$}

The duality between the Heterotic string and F-theory can take two
forms. Starting from the $E_8\times E_8$ theory we have to go through M-theory by
means of a $9-11$-flip from the Heterotic on $T^2$  to Type IA on
$S^1\times S^1/Z_2$ and then by T-duality to Type IIB on $T^2/Z_2$
which in turn is the same as the weak coupling limit of F-theory on
$K3$. For the $Spin(32)/Z_2$ Heterotic theory on $T^2$ we start by
S-duality to Type I and them by two T-dualities to Type IIB on
$T^2/Z_2$ and subsequently to F-theory on $K3$. The gauge fields we
are considering in this sub-section are not charged under the Cartan
of either $E_8\times E_8$ or $SO(32)$.  Therefore, it does not
matter which theory we start from since the mass is the same for the
$SU(3)$ gauge bosons. The authors in
ref\cite{lerche1} considered the $E_8\times E_8$ Heterotic string
since in this case the $E_8\times E_8$ is obtained without any Wilson
line. 

Let us consider the $E_8 \times E_8$ Heterotic string compactified on a torus down to
8 dimensions as in Section 1. We turn off all Wilson lines such that
the moduli that specify the torus can be combined into two complex
scalars  
\bea
U= U_1 + iU_2 &=& \frac{g_{89}}{g_{99}} +i\frac{\sqrt{g}}{g_{99}}
 \nonumber \\
T^0 = T^0_1 + iT^0_2 &=& \f{B}{2} +i \f{\sqrt{g}}{2} \nn
\eea
At the special point in the moduli space of the Heterotic
string where we set the geometric moduli $U =e^{i\pi /3}$ the BPS mass
formula for gauge fields, eq(\ref{8}), is given by\footnote{We set $\alpha_h
' = 2$ in this section.}
\be
M_h^2 = \frac{|(m_8 - U m_9) + T^0(w^9 +U w^8)|}{2T^0_2U_2}
\l{12a}
\ee
We, of course, still have to impose the level matching condition (LM),
eq(\ref{2}), which for states not charged under the Cartan of $E_8\times E_8$
is given by
\be
m_8w^8 + m_9w^9 =1  .
\l{13}
\ee 

If we now approach with the parameter $T^0$ the point $T^0=U$ we have 
\bea
\f{B}{2} + i\f{\sqrt{g}}{2} &=& \f{g_{12}}{g_{22}} + i
\f{\sqrt{g}}{g_{22}} \;\; = \;\; \1/2 + i\f{\sqrt{3}}{2}  \nn \\
&\Rightarrow & B =1 \;\; , \;\; g_{12} =1 \;\; \& \;\;
g_{11}=g_{22}=R^2_c=2 .
\l{14}
\eea
Therefore, we are at the well know point of gauge enhancement of two geometrical $U(1)_G$'s to
$SU(3)_G$. We need six gauge fields to realise this enhancement. The
gauge fields in the Heterotic string together with their quantum
numbers are given in Table 2.
\begin{table}[htb]
\begin{center}
\begin{tabular}{|c|c|c|c|c|c|c|c|c|} \hline
${\bf V}$ & $w^1$ & $w^2$ & $m_1$ & $m_2$ & $p_{R}$ & $p_{L}$ &
$N_{R}$ & $N_{L}$ \\ [2mm] \hline
${\bf V^{1,2}}$& 0 & $\mp 1$ & 0 & $\mp 1$ & 0 & $\pm i \sqrt{2}$ & $\1/2$ & $0$  \\ [2mm]
${\bf V^{3,4}}$& $\mp 1$ & $\pm 1$ & $\mp 1$ & 0 & 0 & $\pm \sqrt{2} \left(
\frac{\sqrt{3}}{2} + \frac{i}{2} \right)$ & $\1/2$ & $0$  \\ [2mm]
${\bf V^{5,6}}$& $\pm 1$ & 0 & $\pm 1$ & $\pm 1$ & 0 & $\pm \sqrt{2} \left(
\frac{\sqrt{3}}{2} - \frac{i}{2} \right)$ & $\1/2$ & $0$  \\ [2mm]
\hline
\end{tabular}
\end{center}
\caption{Additional $SU(3)$ gauge bosons.}
\end{table} 
Note that all states in Table 2 satisfy the level matching condition,
eq(\ref{13}).

Having fixed the value for the quantum numbers identifying the gauge
bosons we can rewrite the formula for their mass, (\ref{12a}), as in Table 3.
\begin{table}[htb]
\begin{center}
\begin{tabular}{|c|c|c|} \hline
${\bf V}$ & $\f{4}{\alpha '} M_h = |P_{R}|$ & LM  \\ [2mm] \hline
${\bf V^{1,2}}$ & $\frac{|T - U|}{2T^0_2U_2} $ &
$2T^0_2U_2 = (U- \bar{T})(\bar{U} - T)$  \\ [2mm]
${\bf V^{3,4}}$& $\frac{|1 +T(U - 1)|}{2T^0_2U_2} $ & $2T^0_2U_2 =|1 +
\bar{T}(U-1)|^2 $    \\ [2mm]
${\bf V^{5,6}}$ & $\frac{|1 +U(T - 1)|}{2T^0_2U_2} $  & $2T^0_2U_2 =
|1 + \bar{U}(T-1)|^2 $  \\ [2mm] \hline
\end{tabular}
\end{center}
\caption{Mass of the $SU(3)$ gauge bosons.}
\end{table} 

Furthermore, using the fact that for $U=e^{i\pi /3}$ we have
\mbox{$\bar{U}=e^{-i\pi /3}=-e^{i\pi-\frac{i\pi}{3}}=-e^{2i\pi /3}=
-U^2$} and $U^2-U+1=0$.  We can rewrite the numerators and
denominators in the expressions in Table 3 as follows
\bea
|1+T(U-1)| &=& |(1-U)(T-U)| \nn \\
|1+U(T-1)| &=& |U(T-U)| \nn \\
|1+T\bar{U}+T| &=& |U^4(U^2 +T)| \nn \\ 
|T-\bar{U}| &=&|T+U^2| \nonumber \\
|1+U(\bar{T}-1)|&=& |U^2(U^2+T)| \nn
\eea
With this result we can write the mass of the 6 $SU(3)$ gauge fields
as in Table 4. All six gauge fields become massless as $T^0\rightarrow U$. 
\begin{table}[htb]
\begin{center}
\begin{tabular}{|c|c|c|} \hline
${\bf V}$ & $\f{4}{\alpha '} M_h = |P_{R}|$ & LM  \\ \hline
${\bf V^{1,2}}$& $\frac{|T - U|}{2T^0_2U_2} $ &
$2T_2U_2 = |(\bar{T} -U)(T + U^2)|$  \\ [2mm]
${\bf V^{3,4}}$& $\frac{|(1-U)(T-U)|}{2T^0_2U_2} $ & 
$2T_2U_2 =|\bar{U}^2(T+U^2)|^2 $ \\ [2mm] 
${\bf V^{5,6}}$ & $\frac{|U(T - U)|}{2T^0_2U_2} $  & $2T_2U_2 =
|\bar{U}(T+U^2)|^2 $   \\ [2mm] \hline
\end{tabular}
\end{center}
\caption{Mass of the additional $SU(3)$ gauge bosons.}
\end{table} 
\\

In F-theory the $K3$ surface we are looking for is the one that has the
gauge group $E_8\times E_8\times U(1)^2$ and such that it depends on
only one moduli to have it enhanced to $SU(3)$. Such a surface is
given by
\be
{\bf K3}: \;\;\;\; y^2+x^3 +fx + g = y^2+x^3+z^5(z-1)(z-z_s)=0
\ee
This surface describes a fibred $K3$ and the fibre has the discriminant given by
\bea
\Delta & = & 4f^3+27g^2 \nn \\
 & = & 27g^2 \nn  \\
 & = & 27z^{10}(z-1)^2(z-z_s)^2
\eea
The points where the discriminant vanishes give the position of
7-branes on $S^2_s$. For this particular expression the gauge group is given by
a fibre $II*$\cite{vafa2} at $0$ and  $\infty$(this point will appear in another patch of
$S^2_s$). And fibre $II$ at $1$ and $z_s$
yielding $E_8\times E_8\times U(1)^2$ symmetry, respectively. The
important point about this $K3$ is that it depends only on one complex
parameter, namely, $z_s$. The point of enhancement to $SU(3)$ is when we
have $z_s \rightarrow 1$ therefore the distance $z_s - 1$ must be
related to the expression $T^0 - U$ in
the heterotic side. 

\begin{figure}[htb]
\epsfysize=5cm
\centerline{\epsffile{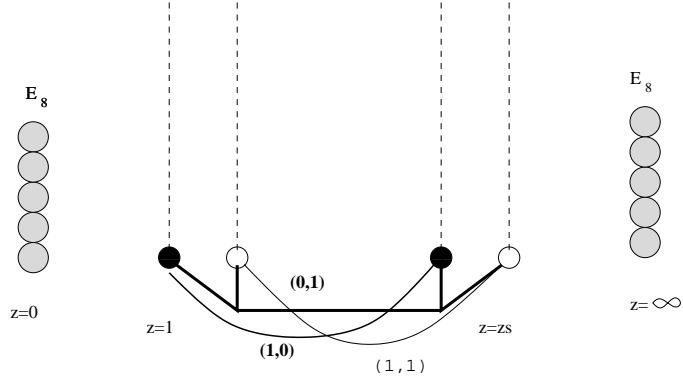}}
\caption{Type IIB gauge bosons enhancing $E_8\times E_8 \times
U(1)^2\rightarrow E_8\times E_8 \times SU(3)$. Each blob represents
one dynamical unit, ${\bf AC}$.}
\end{figure}

In the equivalent Type IIB picture we want to compute explicitly the
mass for a $(p,q)$-string
stretching between the two points of $U(1)$ gauge symmetry, ie, $1$ and $z_s$
(see Fig 2). First a simple analysis of the moduli tells us
that there are 6 strings that can end on the non-local branes sitting at that
points. In our conventions these strings are, up to global $SL(2,Z)$ monodromies,
${\bf V^{1,2}}=\pm (1,0), {\bf V^{3,4}}=\pm (1,1) \,\, \& \,\,{\bf
V^{5,6}}= \pm (0,1)$. This fixes the tension in the mass formula. The
integral in the numerator of the mass formula can be rewritten in terms
of hypergeometric functions after a simple change of variables. We obtain
\be 
{\cal L} = \int_1^{z_s}|\Delta^{-\f{1}{12}}|= \frac{\sqrt{\pi }(-1)^{5/6}(1 - z_s)^{2/3} \Gamma
[5/6]F_{2,1}[5/6,5/6,5/3,1-z_s]}{2^{2/3}\Gamma [4/3]} .
\ee

The periods of  $K3$, using eq(\ref{11}), are also expressible in terms of
hypergeometric functions. In fact,
\bea  
w_0  &=& -2 (-1)^{2/3} \pi {\bf F}_{2,1}[1/6, 1/6, 1, z_s])A(e^{i\pi /3}) \\
w_1 &=& 2(-1)^{5/6}\pi
(-z_s)^{-1/6}F_{2,1}[1/6,1/6,1,\frac{1}{z_s}]A(e^{i\pi /3}) .
\eea

Through a non-trivial hypergeometric transformation we can
rewrite ${\cal L}$ in terms of the periods $w_i$ as follows 
\bea
{\cal L} &=&-\frac{\sqrt{\pi }}{2^{5/3}}\frac{\Gamma
[5/3]\Gamma[1/6]}{\Gamma [4/3]} (e^{-2\pi i /3} w_0 + w_1) \nn \\
&=& -e^{i\pi /3} w_0 + w_1
\eea
Following \cite{lerche1} we divide the original expression for
the mass, eq(\ref{9}), by the fundamental period $w_0$ and use the flat
coordinate for $K3$, $T_{{\tiny IIB}} = w_1/w_0$. The result is
\bea
M_{IIB}(p,q) &=&\biggl|\frac{\eta (e^{i\pi /3})^2}{A(e^{i\pi /3})}\biggr||(p+\tau q)(e^{i\pi /3} - T_{{\tiny IIB}})| \nn  \\
&=& \biggl|\frac{\eta (e^{i\pi /3})^2}{A(e^{i\pi /3})}\biggr||(p+\tau q)(\tau  - T_{{\tiny IIB}})|
\l{14a}
\eea   

Using the $(p,q)$ charges of the strings in Fig. 2 we find complete agreement among the BPS gauge fields
responsible for the enhancement of $E_8\times E_8\times U(1)_G^2
\rightarrow  E_8\times E_8\times SU(3)_G$ in the Heterotic string and
Type IIB by mapping 
\bea
U_h &\rightarrow &\tau  \\
T^0_h &\rightarrow & T_{{\tiny IIB}}
\eea
The overall constants are absorbed in the relation between the metrics of
the two theories.

We now consider the case with Wilson lines turned on.

\subsubsection*{b) An example with non-zero Wilson line: \\
${\bf E_8\times E_8\times SU(3)_G \rightarrow E_8\times
U(1)^2_G\times G_1\times G_2}$}

In Fig 3 we represent the configuration we will
consider in this subsection from the Type IIB perspective. We break one of the
$E_8$'s by moving away an integer number of dynamical units. It is not necessary to
specify in detail what the breaking is at infinity. The elliptic $K3$ surface equivalent to this
configuration in Type IIB is given by
\be
{\bf K3} \;\;\;\;\; y^2 + x^3 + z^5(z-1)(z-z_s)(z-M)^n=0 .
\l{14aa}
\ee
We have put five dynamical units at $z=0$ forming a $E_8$, one
dynamical units at $z=z_s$  and another one at $z=1$
forming $U(1)_G^2$. At $M$ we have a block formed by $n \le 5$ dynamical
units that have moved away from infinity.

\begin{figure}[htb]
\epsfysize=5cm
\centerline{\epsffile{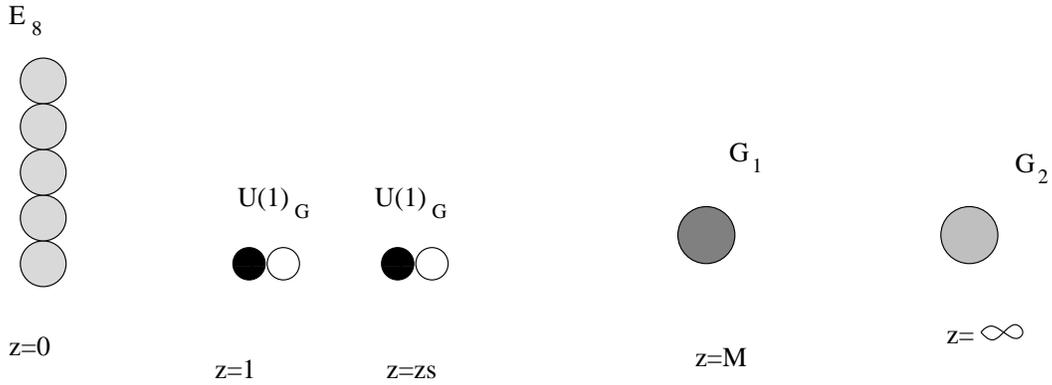}}
\caption{Symmetry breaking. $E_8\times E_8 \times SU(3)_G \rightarrow
E_8\times U(1)^2_G\times G_1\times G_2$. Where $G_1$ and $G_2$ are the
gauge groups left unbroken when $n$ dynamical units move away from
$z=\infty$.} 
\end{figure}

As before we want to compute
\be
{\cal L}=\int_{z_s}^1\Delta ^{-\f{1}{12}} =\int_{z_s}^1z^{-\f{5}{6}}(z-1)^{-\f{1}{6}}(z-z_s)^{-\f{1}{6}}(z-M)^{-\f{n}{6}} .
\l{um}
\ee
For the periods we have similarly
\bea
w_0 &=& \int_{0}^{z_s}
z^{-\f{5}{6}}(z-1)^{-\f{1}{6}}(z-z_s)^{-\f{1}{6}}(z-M)^{-\f{n}{6}} \\
w_1 &=& \int_{0}^{1}
z^{-\f{5}{6}}(z-1)^{-\f{1}{6}}(z-z_s)^{-\f{1}{6}}(z-M)^{-\f{n}{6}}
\l{dois}
\eea

To write ${\cal L}$ in terms of the periods as we did before it turns
out to be convenient this time to Taylor expand ${\cal L}$ and the
periods in a series in $z/M$. We can now write\footnote{We present the details in
Appendix A}  
\be
{\cal L}=\sum_{l=0}^{\infty} C_l(n) \int_{z_s}^1 z^{l-\f{5}{6}}(z-1)^{-\f{1}{6}}(z-z_s)^{-\f{1}{6}} .
\l{tres}
\ee
where $C_l(n)$ are some numerical coefficients. We also obtain similar expressions
for the periods. The point is that we can now
rewrite the integrand in terms of hypergeometric functions and apply
one of Kummer's relations to each element in the sum separately as we
show in appendix A. It turns out that we end up with the following
relation among ${\cal L}$ and the periods
\be
{\cal L} = w_0\tau - w_1 .
\ee
Dividing by $w_0$ we obtain for the mass formula of BPS states stretching between the 7
branes sitting at $z=1$ and those at $z=z_s$
\be
M_{IIB} = \biggl|\f{\eta(e^{\f{i\pi}{3}})^2}{A(e^{\f{i\pi}{3}})}\biggr||(p+\tau
q)(\tau -T_{{\tiny IIB}})|
. \\
\ee
where we introduced once again $T_{{\tiny IIB}}$. Note that $T_{{\tiny
IIB}}$ incorporates all the information on the positions of the branes
as we move them around very much as $T_h$ does with Wilson lines. We
start now to explore how they are connected.

\noindent
{\bf Heterotic String Duals}

Consider a smooth modification of the background of the Heterotic
string considered before by turning on Wilson line moduli. The moduli
now becomes
\bea
T &=& (\f{B}{2} +\f{{\bf A_9 \cdot A_9}}{2}\f{g_{12}}{g_{22}} -
\f{{\bf A_9 \cdot A_8}}{2})  + i \f{\sqrt{g}}{2}(1 + \f{{\bf A_9 \cdot
A_9}}{g_{22}})  \nn \\
&=& T^0 + (\f{{\bf A_9 \cdot A_9}}{2}\f{g_{12}}{g_{22}} -
\f{{\bf A_9 \cdot A_8}}{2})  + i (\f{\sqrt{g}}{2}\f{{\bf A_9 \cdot
A_9}}{g_{22}}) .
\l{14b}
\eea

Let us assume further that the geometrical moduli remain fixed to their
values as in eq(\ref{14}). The BPS mass for the $SU(3)_G$ states is now given by 
\be
M_h = \f{1}{2T^0_2U_2} | m_8 - Um_9 + Tw^9 + (TU + Z)w^8|
\ee

Let us analyse what are the conditions for the $SU(3)_G$ gauge bosons
to become massless again. We surely expect this to be smoothly
related to the Wilson lines parameters. In fact, we have \\

{\bf i)} For ${\bf V^{1,2}}$: 

\be
w^8=0, \;\;  m_8=0,  \;\; m_9 = w^9   .
\ee
We have for the mass of this states  
\be
M_h =\f{1}{2T^0_2U_2} |m_9(U-T)| .
\l{15}
\ee
Level matching condition, eq(\ref{2}), implies that $m_9w^9 =1$, ie $w^9=\pm
1 = m_9$, and this will determine, as before, the $p$ charge of the
$(p,q)$-string in Type IIB. Note that $T$ depends on the Wilson line
parameter now. So to obtain a
massless state we must have, $T\rightarrow U$, or from eq(\ref{14b})  
\bea
{\bf A_9}\cdot {\bf A_9} &=& 0 \rightarrow {\bf A_9} = {\bf 0} \nn \\
\l{16}
{\bf A_9}\cdot {\bf A_8} &=&0 .
\l{17}
\eea 

${\bf ii)}$ For ${\bf V^{3,4}}$ 

\be
m_8 = w^8 = - w^9, \;\;\; m_9 =0
\ee
we obtain for the mass
\bea
M_h &=& \f{1}{2T^0_2U_2}|w^8(1+T - TU -Z)| \nn  \\
&\overrightarrow{=}^{T=U}&  \f{1}{2T^0_2U_2}|w^8(1+U - U^2 -Z)| \nn \\
&\overrightarrow{=}^{U^2-U+1=0}& \f{1}{2T^0_2U_2}|w^8||Z| \;\;\;
\longrightarrow Z=0  .
\l{17a}
\eea  
Note again that level matching requires $w^8= \pm 1= m_8 \rightarrow
w^9 = \mp 1$. Let's analyse the condition on $Z$. Using $U= e^{\f{i\pi
}{3}}$ in eq(\ref{6}) we obtain 
\be
Z= \f{1}{4}({\bf A_9\cdot A_9} +2 {\bf A_9\cdot A_8} - 2{\bf A_8\cdot
A_8}) + i \f{\sqrt{3}}{2}({\bf A_9\cdot A_8} - \f{{\bf A_9\cdot A_9}}{2}) = 0 .
\l{18}
\ee
The imaginary part is equal to zero due to eq(\ref{17}). For the real part we
have 
\be
{\bf A_8\cdot A_8} = 0 \rightarrow {\bf A_8} = {\bf 0} .
\ee
These states become massless only when we turn off the Wilson
lines, ${\bf A_i} ={\bf 0}$,  as expected. Since in this case we are
back to the situation of \cite{lerche1} with the geometrical
parameters fixed to their critical values. And Finally

${\bf iii)}$ For ${\bf V^{5,6}}$:
\be
w^9 = 0, \;\; w^8 = m_8 = m_9
\ee
Level matching requires $w^8 = \pm 1 = m_8= m_9$. Their
mass is given by 
\bea
M_h &\overrightarrow{=}^{T=U}& \f{1}{2T^0_2U_2}|w_8(1-U + TU -Z)| \nn \\
&=& \f{1}{2T^0_2U_2}|w_8||Z| = 0 .
\eea  

So for the geometrical parameters fixed to their critical values,
$B=1$, $R_c^2 = g_{22}=g_{11}=2$ and $g_{12}=1$, we see, by comparing
eq(4.22) with eq(4.26), eq(4.29) and eq(4.33), that the
relative separation parameter of the two dynamical units on Type IIB responsible
for the geometrical enhancement is mapped to the Wilson line moduli by  
\bea
U_h &\rightarrow & \tau  \nn \\
T_h &\rightarrow & T_{IIB}  .
\l{18a}
\eea

If we now consider the more general case when we allow for the geometrical
parameters to be different from their critical values the six gauge
bosons will become massless again when 
\bea
e^{\f{i\pi}{3}} = U &=& T \nn \\
\1/2 + i\f{\sqrt{3}}{2} = \f{g_{12}}{g_{22}} + i \f{\sqrt{g}}{g_{22}}
&=& (\f{B}{2} +\f{{\bf A_9 \cdot A_9}}{2}\f{g_{12}}{g_{22}} - \f{{\bf A_9 \cdot A_8}}{2})  + i \f{\sqrt{g}}{2}(1 + \f{{\bf A_9 \cdot A_9}}{g_{22}})
\l{19}
\eea
If we were to turn on a Wilson line and still keep the $SU(3)_G$
symmetry we would have to tune the geometrical parameters such that the critical values would be shifted as follows
\bea
B &=& 1 + {\bf A_9 \cdot A_8} - \f{{\bf A_9 \cdot A_9}}{2} \\
R_c^2 &=& 2 -\f{{\bf A_9 \cdot A_9}}{2}. 
\eea
Note that the constraints in $Z$, eq(\ref{17a}), would still be in place but now in
the more general form
\bea
{\bf A_9\cdot A_8} &=& \f{{\bf A_9\cdot A_9}}{2} \\ \nonumber
{\bf A_9\cdot A_9} &=& {\bf A_8\cdot A_8} .
\eea

We have shown that by introducing the parameters $U$ and $T$ as in
eq(\ref{5}) and by writing the Heterotic mass formula as in eq(\ref{8}) the map
between Heterotic string BPS states and their duals in Type IIB theory
are immediately obtained, eq(\ref{18a}). In the next section we extend this analysis to
the other non-trivial branch of constant coupling in Type IIB theory.

\subsection{ Branch II - ${\bf \tau =i}$}
For branch II, $f\neq 0 \;\; \& \;\; g= 0 \rightarrow \tau  =i$, and
there are 5 complex degrees of freedom. In fact, $f(\xi )$ is a
polynomial of degree 8 and we have to mod out the $SL(2,{\bf C})$
symmetry. The 24 branes join up forming 8 groups with two mutually
local branes and a non-local one. In our conventions a dynamical
unit is (${\bf AAC}$). In general,  we have  $SU(2)^8$ gauge
group. The other possible gauge groups are

\begin{eqnarray}
\begin{array}{lll}
SU(2)^8   & E_7\times SU(2)^5 & E_7\times E_7\times SU(2)^2\\
SO(8)\times SU(2)^6      & E_7\times SO(8)\times SU(2)^3 & 
                             E_7\times E_7\times SO(8)\\
SO(8)^2\times SU(2)^4    & E_7\times SO(8)^2\times SU(2) &\\
SO(8)^3\times SU(2)^2    & & \\
SO(8)^4                  && \\
\nn
\end{array}
\end{eqnarray}
\smallskip

Therefore, we see that in Branch II we have two basic gauge enhancements
depending on how many dynamical units collide at the same point. We have
\be
(SU(2)\times U(1))^2 \rightarrow SO(8); \;\;\; \& \;\;\; (SU(2)\times
U(1))^3 \rightarrow E_7 \;\;\; ; \nn
\ee

The mass for BPS $(p,q)$-strings in this branch is given by
\be
M_{IIB}(p,q) =\biggl| \f{\eta (i)^2}{A(i)}\biggr| \f{
\int dz |\Delta (z)^{(-1/12)}|}{|\int_{\gamma_0} dz \Delta (z)^{(-1/12)}|}
\ee

Let us consider the map of IIB theory in Branch II to the
$\spin32$ Heterotic String on $T^2$. It is convenient to start with the
following Wilson lines:
\bea
{\bf W_8} &=& (\f{1}{2}^4;\f{1}{2}^4;0^2;0^2;0^4) \nn \\ 
\W9 &=& (0^4;\f{1}{2}^4;0^2;0^2;\f{1}{2}^4) .
\label{21}
\eea
These Wilson lines break the gauge group to $SO(8)^4$. The
semicolon separation will become clear below. 

As part of our duality map we set as before
\be
U_h \rightarrow \tau  .
\ee
In the Heterotic side this implies that
\be
U_h =\f{g_{89}}{g_{99}} + i
\f{\sqrt{g}}{g_{99}} =\tau = i.
\ee
This fixes two of the geometrical moduli,
namely, $g_{89}=0$ and $g_{88}=g_{99} = R^2$. This is why in our redefinition of the
geometrical parameters in Heterotic string with Wilson lines we
left $U_{h}$ unchanged so that this mapping remains the same.

We analyse the map between the Heterotic and type IIB theories in two
examples. Both examples correspond to a one-parameter family in the
respective moduli spaces.

Once we define a one-parameter family in the Heterotic theory we
present a one-parameter family in the Type IIB theory that we argue
is the dual family. There is, obviously, an infinity number of
one-parameter families in a 5 dimensional space as the moduli space we
are dealing with here. Nevertheless, we manage to identify one
potential family in the Type IIB theory. The condition for this family
to be the dual family of the one in the Heterotic theory is that they both have the same
pattern of gauge enhancements taking place in dual points in the
moduli spaces. And the fact that the mass of BPS gauge bosons are the
same every where in moduli space. We verify this to be the case in
both examples we consider. We then write explicitly the duality map
between the moduli of the two theories.  

We start with the enhancement:

\subsubsection*{{\bf a)} ${\bf (SU(2)\times  U(1))^2\times SO(8)^3
\rightarrow E_7^2\times SO(8)}$}

To achieve this enhancement we have to move in the Wilson lines
moduli space by adding to the Wilson lines above the following pieces
\bea
{\bf A_8} &=& {\bf W_8} + (0^4;0^4;a^2,0^2;0^4) \nn \\
{\bf A_9} &=& {\bf W_9} + (0^4;0^4;0^2,a^2;0^4) ,
\eea
where $a$ is a real positive parameter. It is immediate to see that
the effect of the perturbation is to break the third $SO(8)_{(3)}$, ie,
the one occupying the third block of four slots in the lattice vector above, to
$(SU(2)\times U(1))_{(1)}\times (SU(2)\times U(1))_{(2)}$. We
anticipate the map with type IIB by picturing this breaking in Fig 4.

\begin{figure}[htb]
\epsfysize=5cm
\centerline{\epsffile{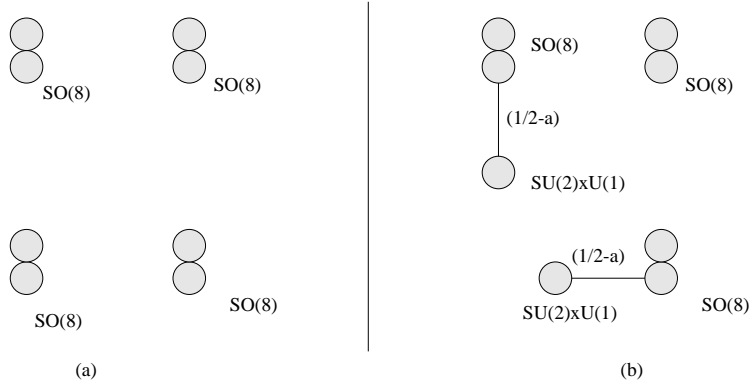}}
\caption{Symmetry breaking in Branch II. Each blob represents one
dynamical unit ${\bf AAC}$. a) $SO(8)^4$ b)$(SU(2)\times
U(1))_{(1)}\times (SU(2)\times U(1))_{(2)}\times SO(8)^3$}
\end{figure}

Note that for $a = \frac{1}{2}$ we have $(SU(2)\times U(1))_{(1)} \rightarrow
SO(4)_{(1)}$ and similarly  $(SU(2)\times U(1))_{(2)} \rightarrow
SO(4)_{(2)}$ since the following vectors become massless (we will
concentrate on one of the enhancements to $E_7$ only since the process is
identical for both)
\be
{\bf Q}^{(1)} = \pm (0^4;0^4;+1,+1;0^2;0^4) .
\l{23}
\ee
We also have states in the vector representation of $SO(4)_{(1)}
\times SO(8)_{(1)}$ becoming massless when $a=1/2$\footnote{Here $p$
stands for all permutations within the bracket} 
\be
{\bf Q}^{(2)}=(\underbrace{ \pm 1,0^3}_{\mbox{p}};0^4;\underbrace{ \pm
1,0}_{\mbox{p}};0^2;0^4) = (8_v;1,4;1;1) .
\l{24}
\ee
These gauge bosons enhance $(SO(4)_{(1)}\times SO(8)_{(1)})
\rightarrow SO(12)_{(1)}$. Analogously, we also have
$(SO(4)_{(2)}\times SO(8)_{(4)}) \rightarrow SO(12)_{(2)}$.  The mass
of this states as they approach $a=\1/2$ are listed in Table 5.

If this model is to represent a configuration of branes in Type IIB in
Branch II there can not be a $SO(12)$. But we are not finished yet. We
have to check the states with winding numbers.
 
If we rewrite the expression for the heterotic momenta using
$U_h = i$ we can write for the mass\footnote{This is basically two
copies of the 9 dimension case reviewed in \cite{gab3}.}, eq(\ref{1a}),
\be
M_h^2 = P_{iR}^2 = (p_8 - (g_{88}w^8 - Bw^9)\frac{1}{\alpha '})^2g^{88}
+ (p_9 - (g_{99}w^9 + Bw^8)\frac{1}{\alpha '})^2g^{99}  
\ee
and  using $g^{88} = g_{99}/g = 1/g_{88} = R^{-2} = g^{99}$ we write
\be
M_h^2 = P_{iR}^2 = (p_8 - (R^2w^8 - Bw^9)\f{1}{\alpha '})^2R^{-2}
+ (p_9 - (R^2w^9 + Bw^8)\frac{1}{\alpha '})^2R^{-2}  
\ee  
For massless states we must have
\bea
p_8 &=& ((R^2w^8 - Bw^9)\f{1}{\alpha '}) \label{24a} \\
p_9 &=& (R^2w^9 + Bw^8)\frac{1}{\alpha '})
\eea

Substituting this in the formula for level matching condition for
gauge bosons, $P_L^2 =4/\alpha ' $, we obtain
\be
R^2((w^8)^2 + (w^9)^2) = \alpha ' (1 - \f{q^2}{2})  
\ee
This equation requires that we have $q^2 \le 2$. Defining $\Lambda = q
+ \lambda$, $\lambda \in \Gamma^{16}$ such that $\Lambda^2$ is
smallest. We obtain
\be
R^2((w^8)^2 +(w^9)^2) = \alpha ' (1 - \f{\Lambda ^2}{2})  
\ee

We separate the winding states in odd and even values for the sake of
the analysis. Furthermore, it is easy to see that for odd winding numbers
only $w^i=\pm 1$ contribute with massless states. For $w^9 =\pm 1$ and
$w^8=0$ we obtain ${\bf q} = {\bf Q} \pm
(0^4;\f{1}{2}^{4};0^2;(\1/2)^6)$. Choosing\footnote{Where (e \# -)
stands for an even number of minus signs in the permutations.}
\be
{\bf Q}= \mp (\underbrace{\underbrace{(\pm
\f{1}{2})^4};(\1/2)^4;\underbrace{(\pm \1/2)^2}}_{\mbox{(e\# -)}};(\1/2)^2;(\1/2)^4)
\l{25}
\ee
we obtain
\be
\Lambda = (\underbrace{(\pm \f{1}{2})^4;0^4;(\pm \1/2)^2}_{\mbox{(e\#
-)}};0^2;0^4),
\l{26}
\ee
These are the spinors ${\bf 32}_{\pm 1}$ of $SO(12)_{(1)}$. The subscript is the
$U(1)_{w^9}$ charge associated to winding along
$x^9$-direction. This gauge vectors become massless at the critical
radius $R_c^2 = \f{\alpha '}{4}$. It turns out that it is convenient to
write the spinors in terms of the gauge groups $SO(8)_{(1)} \times
(SU(2)\times U(1))_{(1)}$, ie, ${\bf 32}_{\pm 1} = ({\bf 8_s};{\bf 1};{\bf
1};{\bf 1};{\bf 1})_{\pm \1/2} + ({\bf 8_c};{\bf 1};{\bf
2};{\bf 1};{\bf 1})$. For even winding numbers, only $w^i=\pm 2$
contribute. In this case we take ${\bf Q} = \pm (0^4;(1)^{4};0^2;(1)^6)$ and
$\Lambda^2=0$, these are singlets, $({\bf 1,1,1,1,1})_{\pm 2}$. These singlets
become massless at the same radius as the spinors. So for $R_c^2
=\f{\alpha '}{4}$ the spinors and singlets enhance  $SO(12)_{(1)}\times U(1)_{w^9}$ to
$E_7^{(1)}$. Analogously, gauge fields with quantum numbers $w^9=0$
and $w^8=\pm 1,\pm 2$ enhance $SO(12)_{(2)} \times U(1)_{w^8}$ to $E_7^{(2)}$. 

Note that from eq(\ref{24a}) when $w^8=0$ and $w^9 = \pm 1,\pm 2$, as
above, we have a constraint in the value of the anti-symmetric field
$B$. A straight forward analysis\footnote{In Appendix B we present
some details.} of that equation shows that we must
have $B \in 2Z$ (this only fixes $B$ up
to an $SL(2,Z)$ transformation). Therefore, we choose $B=0$. This
fixes the real part in the mass formula, eq(\ref{8}), as in Table 5.

Consider now the other two free moduli. We have the Wilson line parameter,
$a$, and the radius, $R^2= g_{88}=g_{99}$. As mentioned before, we
will consider only one-parameter families. So we have to fix one
of this parameters. We know that when we are at the critical radius as
$a\rightarrow 1/2$ we obtain the enhancements described
above. Therefore, we choose the radius to be a function of $a$ so
that at $a=1/2$ we have $R^2 =R_c^2= \alpha ' /4$. Of course
there are an infinity number of ways of achieving this but ultimately
we want a match between the masses of BPS states in the dual
theories. The analysis carried out in Appendix B determines that we
set for the specific choice of parameters in the Type IIB side to
discussed below
\be
R^2 =  \alpha ' \bigl(a(1-a) - (\1/2 -a)\bigr)  .
\ee

It is now possible to  write the mass formula for all the BPS states we have
been considering in this section. We write their mass
in terms of the mass formula, eq(\ref{8}). The results are summarised
in Table 5. 
\begin{table}[htb]
\begin{center}
\begin{tabular}{|c|c|} \hline
${\bf Q}$ & $M_h$ \\ \hline
$(\pm (0^4;0^4;+1,+1;0^2;0^4)$ & $\f{4}{\alpha '}|2(\f{1}{2} -a)|$ \\
$({\bf 8_v};{\bf 1};{\bf 2};{\bf 1};{\bf 1})_{0}$ & $\f{4}{\alpha '}|(\f{1}{2} -a)|$ \\
$({\bf 8_v};{\bf 1};{\bf 1};{\bf 1};{\bf 1})_{\pm 1}$ & $\f{4}{\alpha '}|(\f{1}{2} -a)|$ \\
$({\bf 8_c};{\bf 1};{\bf 2};{\bf 1};{\bf 1})_{0}$ & $\f{4}{\alpha '}| i(\f{1}{2} -a)|$ \\
$({\bf 8_s};{\bf 1};{\bf 1};{\bf 1};{\bf 1})_{\pm \1/2}$ & $\f{4}{\alpha '}|(-1+i)(\f{1}{2} -a)|$ \\
$({\bf 1};{\bf 1};{\bf 1};{\bf 1};{\bf 1})_{\pm 2} $ &$\f{4}{\alpha
     '}|(2 i(\f{1}{2} -a)|$ \\ \hline
\end{tabular}
\end{center}
\caption{Additional gauge bosons for $SO(8)_{(1)}\times(SU(2)\times
U(1))_{(1)}\times U(1)_{w^9}\rightarrow E_7^{(1)}$}
\end{table}

\noindent 
{\bf Type IIB duals}

In type IIB theory starting from an $SO(8)^4$ configuration we move
the two dynamical units forming one of the $SO(8)$s in perpendicular
directions toward two $SO(8)$s (see Fig. 4). As each of the dynamical
units approach one of the $SO(8)$s a number of $(p,q)$-strings will
become massless. We identify this strings by their symmetry properties
and compute their mass near the point they become massless.

For the $SU(2)\times U(1)$ in  Type IIB we have ${\bf AAC}={\bf
BAA}$. The two vectors in the adjoint of $SU(2)$ correspond to the
${\bf A}\rightarrow {\bf A}$ $(1,0)$-string with both
orientations. The enhancement to $SO(4)$ occurs when strings (prongs)
like the ones in Fig. 6 become massless. They are equivalent to the
vectors ${\bf Q}^{(1)}= \pm (0^4;0^4;+1,+1;0^2;0^4)$. The $\pm$
correspond to orientation.  

\begin{figure}[htb]
\epsfysize=5cm
\centerline{\epsffile{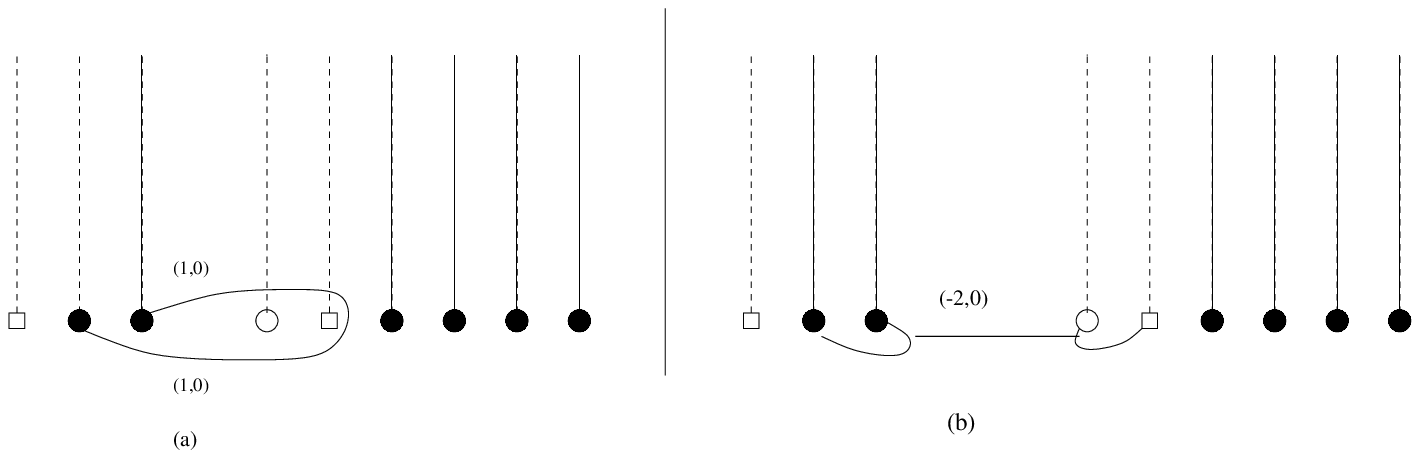}}
\caption{For ${\bf Q^{(1)}}=\pm (0^4;0^4;1,1;0^2;0^4)$.  In b) we show the string junction
equivalent to the string going across branch cuts in a). They are
connected through a string creation process.}
\end{figure}

For the states in the vector representation, ${\bf Q}^{(2)}
 =({\bf 8_v},{\bf 1},{\bf 4};{\bf 1};{\bf 1})$, of $SO(8)_{(1)}\times
 SO(4)_{(1)}$ we separate it in two parts in terms of
 $SO(8)_{(1)}\times (SU(2)\times U(1)^2)_{(1)}$. One where the
 permutations have the same signs, ${\bf Q}_a^{(2)} = (\pm
 \underbrace{+1;0^3}_{\mbox{p}};0^4;\underbrace{+1,0}_{\mbox{p}};0^2;0^4)$ and the other where the permutations have opposite signs, ${\bf Q}_b^{(2)} =\pm
(\underbrace{-1;0^3}_{\mbox{p}};0^4;\underbrace{+1,0}_{\mbox{p}};0^2;0^4)$.
 We identify ${\bf Q}_a^{(2)}$ with the strings (prongs) in Fig 6 and
 ${\bf Q}_b^{(2)}$ with Fig 7.

\begin{figure}[htb]
\epsfysize=5cm
\centerline{\epsffile{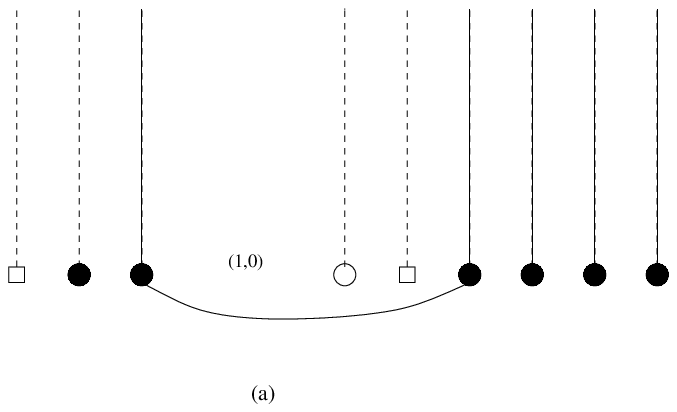}}
\caption{${\bf Q}_a^{(2)}$ }
\end{figure}

\begin{figure}[htb]
\epsfysize=5cm
\centerline{\epsffile{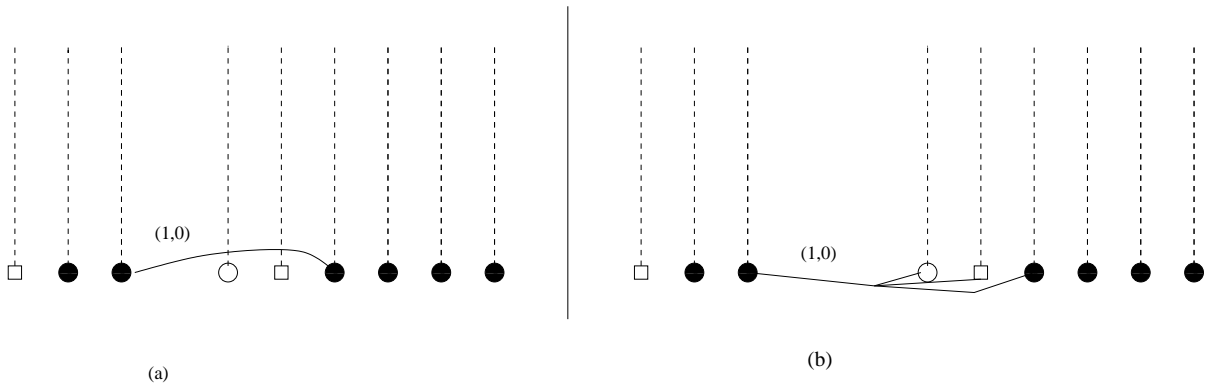}}
\caption{For ${\bf Q}_b^{(2)}$. Two equivalent configurations are shown.}
\end{figure}

For the spinors we identify ${\bf q}_a= ({\bf 8_c};{\bf 1};{\bf 2};{\bf
1};{\bf 1})$ with the configurations exemplified in Fig 8. And ${\bf
q}_b= ({\bf 8_c};{\bf 1};{\bf 2};{\bf 1};{\bf 1})$ is identified with Fig 9. 

\begin{figure}[htb]
\epsfysize=5cm
\centerline{\epsffile{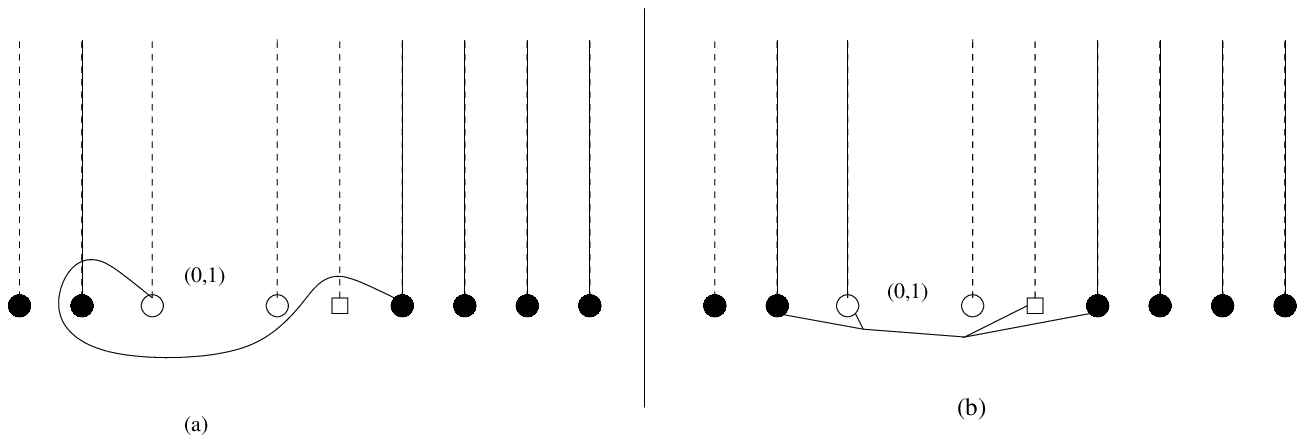}}
\caption{For ${\bf q}_a =({\bf 8_c};{\bf 1};{\bf 2};{\bf
1};{\bf 1})$. Two equivalent representations of this BPS state are
shown in a) and b).}
\end{figure}

\begin{figure}[htb]
\epsfysize=5cm
\centerline{\epsffile{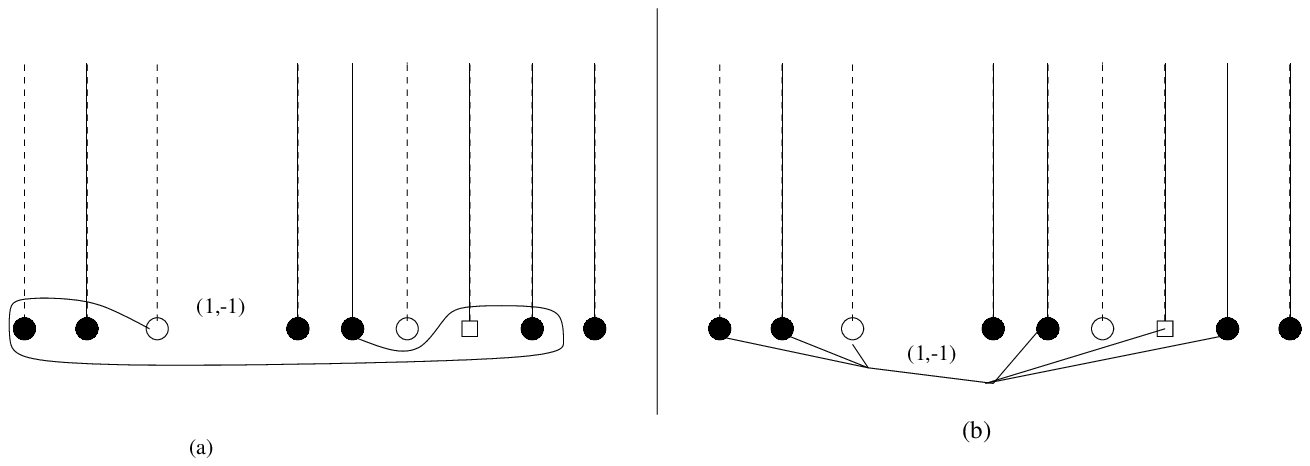}}
\caption{For ${\bf q}_b= ({\bf 8_c};{\bf 1};{\bf 2};{\bf
1};{\bf 1})$. The O7-plane ({\bf CB}) branch cut has been shifted to
the right to simplify the drawing. Once again two equivalent
configurations are shown.}
\end{figure}

We now compute the mass for each of this configurations. The $K3$
surface equivalent to the configuration of branes in Fig. 4  is given by
\be
{\bf K3} \;\;\; : \;\;\; y^2 + x^3 + x(z^{2}(z-1)^{2}(z-z_s))
\ee
We obtain for the integral of the discriminant of this $K3$
\be
{\cal L} = \int_1^{z_s} \Delta^{-\f{1}{12}}    = i\sqrt{\pi
}(1-z_s)^{\f{1}{4}}\f{\Gamma [\f{3}{4}]}{\Gamma [\f{5}{4}]} F_1(\1/2 ,
\1/2 , \f{5}{4}, 1-z_s)
\ee

For the periods we obtain
\bea
w_0 &=& - (-1)^{\f{1}{4}}\sqrt{\pi
}(z_s)^{\f{1}{4}}\f{\Gamma [\f{3}{4}]}{\Gamma [\f{5}{4}]} F_1(\1/2 ,
\1/2 , \f{5}{4},z_s) \\
w_1 &=& - i (-1)^{\f{1}{4}}\pi (z_s)^{-\f{1}{4}} F_1(\f{1}{4} ,
\1/2 , 1,\f{1}{z_s}) .
\eea
By means of a non-trivial Hypergeometric transformation we can write
\be
{\cal L} = w_0 + i w_1
\ee
And we obtain for the mass of a $(p,q)$-string 
\be
M_{IIB}(p,q)= \biggl|\f{\eta (i)^2}{A(i)}\biggr||(p + iq)(i)(i-T_{{\tiny IIB}})|
\ee
We can now compare the masses of BPS gauge fields in both
theories. The results are summarised in Table 6. It is clear that the
explicit duality map is given by
\bea
U_h &\rightarrow & \tau   \nn \\
|(\1/2 -a)|&\rightarrow &|i(i-T_{{\tiny IIB}})|   .
\l{map4}
\eea 

\begin{table}[htb]
\begin{center}
\begin{tabular}{|c|c|c|c|c|c|} \hline
${\bf V}$ & $\f{\alpha '}{4} M_h$ & $(\f{A(i)}{\eta (i)^2})M_{IIB}$ \\ \hline
$(\pm (0^4;0^4;+1,+1;0^2;0^4)$ & $|(2)(\f{1}{2} -a)|$ &
$|(2)[i(i-T_{{\tiny IIB}})]|$ \\
$({\bf 8_v};{\bf 1};{\bf 2};{\bf 1};{\bf 1})_{0}$ & $|(\f{1}{2} -a)|$ & $|[i(i-T_{{\tiny IIB}})]|$ \\
$({\bf 8_c};{\bf 1};{\bf 2};{\bf 1};{\bf 1})_{0}$ & $|(i)(\f{1}{2} -a)|$ & $|(i)[i(i-T_{{\tiny IIB}})]|$ \\
$({\bf 8_s};{\bf 1};{\bf 1};{\bf 1};{\bf 1})_{\pm \1/2}$ & $|(-1 + i)(\f{1}{2} -a)|$ & $|(-1+i)[i(i-T_{{\tiny IIB}})]|$ \\ 
$({\bf 1};{\bf 1};{\bf 1};{\bf 1};{\bf 1})_{\pm 2}$ & $|(2i)(\f{1}{2} -a)|$ & $|(2i)[i(i-T_{{\tiny IIB}})]|$  
\\ \hline
\end{tabular}
\end{center}
\caption{Heterotic-IIB map for the gauge bosons of $SO(8)_{(1)}\times(SU(2)\times
U(1))_{(1)}\times U(1)_{w^9}\rightarrow E_7^{(1)}$.}
\end{table}

In the next sub-section we consider the vector bosons responsible for
the enhancement of $SU(2)\times U(1) \rightarrow SO(8)$.

\subsubsection*{{\bf b)} ${\bf SO(8) \rightarrow SU(2)\times U(1)}$}

In the previous section we identified the $(p,q)$-strings
configurations that are responsible for the enhancement to $E_7$. To
complete the symmetry enhancements possible in the this branch we here
consider the enhancement $(SU(2)\times U(1))^2 \rightarrow SO(8)$. We
choose, for convenience, to do this through  $E_7^2\times (SU(2)\times
U(1))\times (SU(2)\times U(1)) \rightarrow E_7^2 \times SO(8)$ (see
Fig. 10). We choose to put one of the $E_7$'s at $z = \infty$ and the other one
at $z=0$. We also put one of the $SU(2)\times U(1)$ at $z=1$ and the
other at $z=z_s$. 
\begin{figure}[htb]
\epsfysize=5cm
\centerline{\epsffile{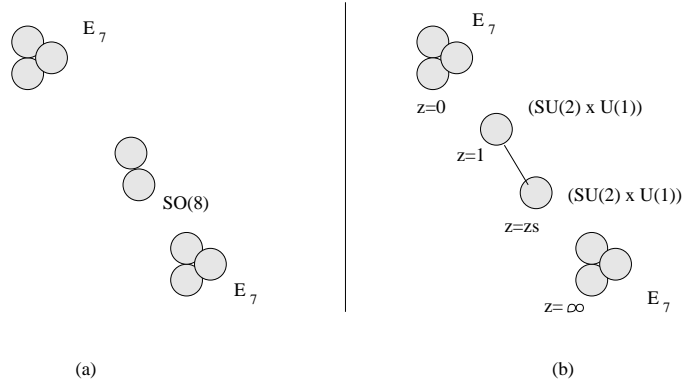}}
\caption{$E_7^2\times (SU(2)\times
U(1))\times (SU(2)\times U(1)) \rightarrow E_7^2 \times SO(8)$}
\end{figure}

The mass of a $(p,q)$-string stretching from one dynamical unit (${\bf
AAC}$) to the other is of the form
\be
M(p,q) = \biggl|\frac{\eta (i)^2}{A(i)}\biggr||(p + iq)\frac{{\cal L}}{w_0}|
\ee
With the configuration above we obtain for the integral
of the discriminant
\be 
{\cal L} = \f{2}{\sqrt{\pi}} e^{3\pi
i/4}(1-z_s)^{1/2}\Gamma [3/4]^2 F{1,2}(3/4,3/4,3/2,1-z_s)
\ee
For the $K3$ periods we obtain
\bea
w_0 &=& e^{i\pi /2} \sqrt{2}\pi F_{1,2}(1/4,1/4,1,z_s) .\\
w_1 &=& e^{- i \pi /4} \sqrt{2} \pi
(-z_s^{-1/4})F_{1,2}(1/4,1/4,1,1/z_s).
\eea
We now use one of Kummer's relations for Hypergeommetric
functions to write
\be
\f{\sqrt{\pi}}{2\Gamma [3/2]\Gamma [3/4]} {\cal L} = \f{\Gamma
[1/4]}{\sqrt{2}\pi} e^{i\pi /2}w_0 - \f{\Gamma [1/4]}{\sqrt{2}\pi }w_1
.
\ee
Therefore we have for the mass of BPS $(p,q)$-strings in this background
\be
M_{BPS}^{IIB}(p,q) = \biggl|\f{\eta (i)}{A(i)}\biggr||(p + iq)(i-T_{{\tiny IIB}} )| .
\ee

In the heterotic side we have the vectors below becoming massless
to form the ${\bf Adj}$ of $SO(8)$. In the respective figures we draw
the $(p,q)$-strings we identify as the dual pairs in the IIB side.
\bea
{\bf Q}^{(1)} &=& \pm (0^4,+1,+1,0^2;0^8)\;\; \mbox{or} \;\; \pm
(0^4;0^2;+1,+1;0^8) \rightarrow \mbox{Fig. 12}  \l{27} \\
{\bf Q}^{(2)} &=& \pm (0^4;-1,0;+1,0;0^8) \rightarrow \mbox{Fig. 13}
\l{28} \\
{\bf Q}^{(3)} &=& \pm (0^4;+1,0;+1,0;0^8) \rightarrow \mbox{Fig. 14} \l{29}
\eea

In Table 7 we list the masses and quantum numbers of this gauge
fields in both theories. The duality map for the moduli space is given
by
\bea
U_h & \rightarrow & \tau  \nn \\
|(1+i)(\1/2 -a)| &\rightarrow & |i-T_{{\tiny IIB}} | .
\l{map5}
\eea
\begin{table}[hbt]
\begin{center}
\begin{tabular}{|c|c|c|c|c|c|c|c|} \hline
 $Q $ & $m$ & $\f{\alpha '}{4}M_h$ & $p$ & $q$ & $\f{A(i)}{\eta (i)^2}M_{IIB}$ \\ \hline
$\pm (0^4,+1,+1,0^2;0^8)$ & $\pm 1$ &$|(1-i)[(1+i)(\f{1}{2} -a)]|$&$-1$ &$1$ & $|(1-i)(i-T_{{\tiny IIB}} )|$ \\
$\pm (0^4,0^2,+1,+1;0^8)$ & $\pm 1$ &$|(1-i)[(1+i)(\f{1}{2} -a)]| $&$-1$ &$1$ & $|(1-i)(i-T_{{\tiny IIB}} )|$\\
$\pm (0^4,+1,0,+1,0;0^8)$ & $\mp 1$ &$|(i)[(1+i)(\f{1}{2} -a)]|$&$0$ &$1$ & $|(i)(i-T_{{\tiny IIB}} )|$ \\
$\pm (0^4,-1,0,+1,0;0^8)$ & $0$ &$|[(1+i)(\f{1}{2} -a)]|$&$1$ &$0$ & $|(i-T_{{\tiny IIB}} )|$ \\ \hline
\end{tabular}
\caption{BPS gauge fields enhancing $(SU(2)\times U(1)) \times
(SU(2)\times U(1)) \rightarrow SO(8)$ in the Heterotic and IIB. Also
$m_8=m_9=m$ and $w^8=w^9=w=0$.}
\end{center}
\end{table}

\begin{figure}[htb]
\epsfysize=5cm
\centerline{\epsffile{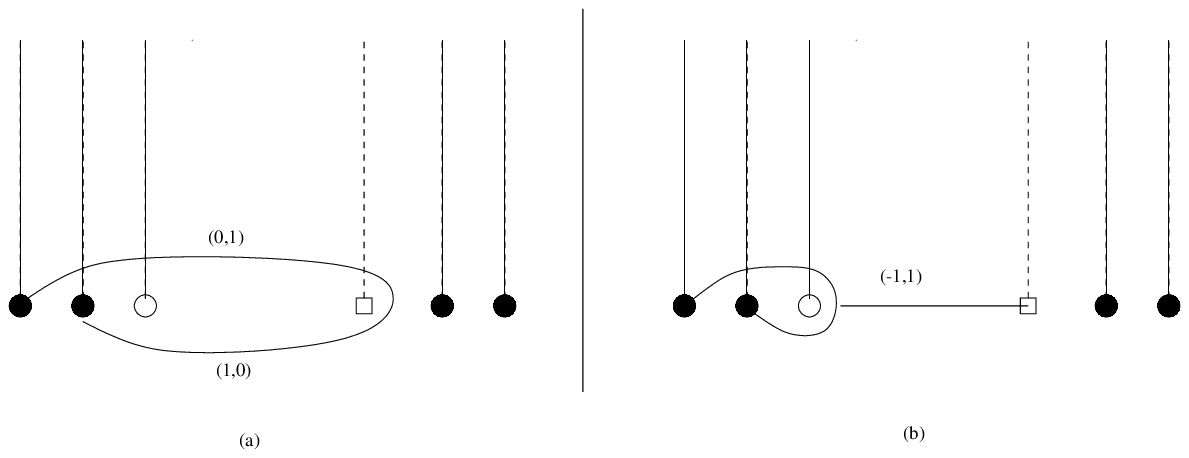}}
\caption{${\bf Q}^{(1)}$ for $(SU(2)\times U(1)) \times
(SU(2)\times U(1)) \rightarrow SO(8)$. Two equivalent configurations
are shown. }
\end{figure}

\begin{figure}[htb]
\epsfysize=5cm
\centerline{\epsffile{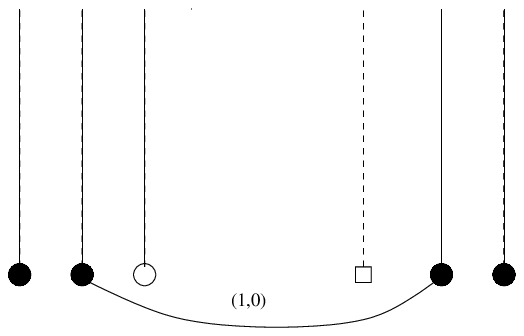}}
\caption{${\bf Q}^{(2)}$ for $(SU(2)\times U(1)) \times
(SU(2)\times U(1)) \rightarrow SO(8)$.}
\end{figure}

\begin{figure}[htb]
\epsfysize=5cm
\centerline{\epsffile{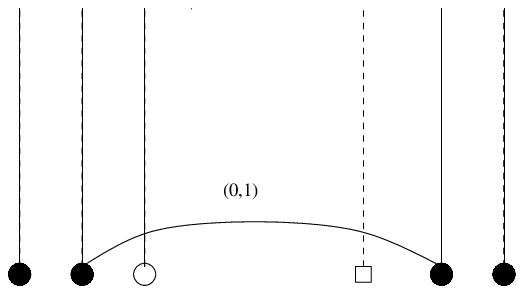}}
\caption{${\bf Q}^{(3)}$ for $(SU(2)\times U(1)) \times
(SU(2)\times U(1)) \rightarrow SO(8)$.}
\end{figure}

\section{Conclusion}

In this paper we have analysed the duality between the Heterotic
string in 8 dimensions and Type IIB theory in the base space of an
elliptically fibred $K3$. The duality map between the
relevant BPS states was given in detail in the branches of moduli space of Type IIB with
constant coupling. 

In Type IIB the flat coordinate, $T_{IIB}=w_0/w_1$, defined in
terms of the periods of the underlying $K3$ geometry encompasses all
the information on the position of the 7-branes in the background
when the elliptic fibre has constant complex structure. It is
therefore the natural coordinate to be used when analysing the map
with Wilson lines on the Heterotic string side. In fact, redefining
the Kahler structure of the torus, $T_h$, in the Heterotic
theory to include information on the Wilson lines. We have found that
it becomes the natural coordinate to account for the effects of Wilson
lines to the mass of BPS states.

For the the case of the $Spin(32)/Z_2$ Heterotic string the masses
of BPS states for specific values of the Wilson lines were analysed in
detail. In particular, we considered the enhancements: $SO(8)\times SU(2)\times U(1)\times
U(1)_G \rightarrow E_7$ and $(SU(2)\times U(1))^2 \times SO(8)$. In both
cases we identified the dual BPS string junctions responsible for the
enhancement on the Type IIB side and computed their masses finding
complete agreement between the two sets of states.

It would be interesting to consider the case with non constant
coupling in Type IIB. Work in this direction is under way.

\section*{Acknowledgements}
I would like to thank Matthias Gaberdiel for explaining to me many
aspects of his works and for suggestions. I thank also Michael Green
for his continuous support and valuable comments and explanations. I
would also like to thank W. Lerche for correspondence. I have enjoyed
very helpful conversations with Fernando
Quevedo, Tathagata Dasgupta and Pierre Vanhove. 

I would like to acknowledge financial support from CNPq (Brazilian Ministry of Science) through a
PhD. Scholarship. I also acknowledge partial financial support from the
Cambridge Overseas Trust.

\appendix
\section{Generalising The Hypergeometric Relation Among ${\cal L}$,
$w_0$ and $w_1$.}
\setcounter{equation}{0}

Several times in this paper we had to rewrite the distance ${\cal  L}$
in terms of the periods of $K3$. In most cases when we have only four
groups of dynamical branes in $S^2_s$ we can rewrite the integrals in
terms of hypergeometric functions. It turns out then that the
integrals in this form are related by means of the Kummer's
relations\cite{kummer}. However, when the blocks break apart and we
have more then four points in the sphere we cannot write then as
hypergeometric functions anymore. Nevertheless, we show that by Taylor
expanding the integrand we can write the integrals as a sum of
integrals that in turn can be related to
hypergeometric functions. We can then apply Kummer's relations to each
element of the sum in separate and sum up the series again to obtain the
relation we look for. This generalises the hypergeometric relations to
a more general set of integrals. In this appendix we present the details of this
calculations for the configuration considered in Section 4.1.

We start with the length of the string stretched from $z=1$ and
$z=z_s$
\be
\int_{z_s}^1\Delta ^{-\f{1}{12}} =\int_{z_s}^1
z^{-\f{5}{6}}(z-1)^{-\f{1}{6}}(z-z_s)^{-\f{1}{6}}(z-M)^{-\f{n}{6}} .
\l{uma}
\ee
we now Taylor expand $(z-M)^{-\f{n}{6}}$ as
\be
(z-M)^{-\f{n}{6}} = \sum_{l=0}^{\infty}C_l(n)z^l
\l{umaa}
\ee
where $C_l(n)$ are standard numerical coefficients. If we now plug this back in eq(\ref{uma}) we have
\be
{\cal L} = \sum_{l=0}^{\infty} C_l(n)  \int_{z_s}^1  z^{l-\f{5}{6}}(z-1)^{-\f{1}{6}}(z-z_s)^{-\f{1}{6}} .
\l{tresa}
\ee
This integral can be written in terms of hypergeometric functions by
doing a simple change of variables, ie, $z\rightarrow
(z-z_s)/(1-z_s)$. We obtain
\be
{\cal L} =\sum_{l=0}^{\infty} C_l(n)\biggl(\f{\sqrt{\pi
}(-1)^{5/6}}{2^{2/3}}\f{\Gamma [\f{5}{6}]}{\Gamma[\f{4}{3}]}
F_{2,1}[\f{5}{6},(\f{5}{6}-l),\f{5}{3}, 1-z_s]\biggr) .
\l{umaaa}
\ee
Similarly, we obtain for the periods, eq(4.17) and eq(4.18),
\bea  
w_0 &=& \sum_{l=0}^{\infty} C_l(n)\biggl(\f{(1-i\sqrt{3})}{2}\f{\Gamma
[\f{5}{6}]\Gamma[ \f{1}{6} +l]}{\Gamma[1+l]}z_s
F_{2,1}[\f{1}{6},(\f{1}{6}+l),(1+l),z_s]\biggr) \\
w_1 &=& -\sum_{l=0}^{\infty} C_l(n)\biggl((-1)^{5/6}\f{\Gamma[\f{5}{6}]\Gamma[\f{1}{6} +l]}{\Gamma[1+l]}(-z_s)^{-\f{1}{6}}
F_{2,1}[\f{1}{6},(\f{1}{6}+l),(1+l), \f{1}{z_s}]\biggr) .
\eea

Using the following Kummer relation\cite{kummer}
\bea
e^{i\pi 5/6}\f{ \Gamma[\f{5}{6}] \Gamma[\f{5}{6}]}{
\Gamma[\f{5}{3}]}(1-z_s)^{2/3}F_{2,1}[\f{5}{6},\f{5}{6}-l,\f{5}{3},1-z_s]
&=& \f{ \Gamma[\f{5}{6}] \Gamma[\f{1}{6} + l]}{
\Gamma[1+l]}z_sF_{2,1}[\f{1}{6},\f{1}{6}+l,1+l,z_s] + \nn \\
&=& e^{i\pi 5/6}\f{ \Gamma[\f{5}{6}] \Gamma[\f{1}{6} + l]}{
\Gamma[1+l]}(-z_s)^{-\f{1}{6}}F_{2,1}[\f{1}{6},\f{1}{6}+l,1+l,\f{1}{z_s}]
. \nn
\eea
we can write each element in the series representation of ${\cal L}$
in terms of the respective elements in the series representation of
$w_0$ and $w_1$. Plugging this result back in the sum, eq(\ref{umaaa}), and reexpressing
the sum in its closed form, we arrive at the desired relation among
${\cal L}$ and $w_0$ and $w_1$, ie,
\be
{\cal L} = \tau w_0 - w_1 .
\ee

It is clear that this procedure can be applied to any distribution of
the dynamical units on the sphere by Taylor expanding an appropriate
number of terms in the expression for ${\cal L}$, $w_0$ and $w_1$.  

\section{Fixing Moduli in Branch II}
\setcounter{equation}{0}

In Section 4.2 the first example of gauge enhancement in Branch II to
be analysed was
$SO(8)^3\times (SU(2)\times U(1))^2 \rightarrow E_7^2\times SO(8)$. To
obtain this enhancement the following Wilson lines were turned on
\bea
A_8 &=& (\1/2^4,\1/2^4,a,a,0^2,0^4)  \\
A_9 &=& (0^4,\1/2^4,0^2,a,a,\1/2^4) 
\eea
with $a=0$ and $a=1/2$ being the critical values for the Wilson lines
parameter. This was enough to determine the masses of BPS gauge bosons
transforming in the vector and adjoint representations of the gauge
groups. With this information we fixed part of the map with the BPS
states on the Type IIB theory. We saw also that to obtain the full
gauge enhancement we needed BPS states transforming in the spinor
and singlet representations of the gauge groups. However, for these
states to become massless we need to tune not only the parameter in
the Wilson lines but the geometric moduli as well. We want to
determine the value of the geometric moduli, $R$ and $B$,
such that the masses of the BPS states responsible for the enhancement
above match with those in Type IIB. We will concentrate in three
specific examples of gauge fields here. The
quantum numbers of these states will be given explicitly. The results
easily generalise to all other states.

We will concentrate on the spinor with quantum numbers $w^9=+1$
and $w^8=0$. And the singlet with quantum numbers, $w^9=+2$
and $w^8=0$. The other quantum numbers will be determined below.

Recall that the critical radius was, eq(4.53), determined in terms of \footnote{In this section $\alpha_h' = 2$.}
\be
\Lambda = (\pm\1/2^4, 0^4,\pm\1/2^2,a-\1/2,a-\1/2,0^4)
\ee
as $R_c^2/2 =1-\Lambda^2/2= a(1-a)$. Was the
radius to be set in this form we would have the spinors and singlets all
massless for all values of $a$. However this would generate
enhancements that have no equivalent in Branch II on Type
IIB. But we know that when $a=1/2$ we must have the appropriate
enhancement. One way to guarantee that this is the case is to set
$R^2/2=1/4$. But it turns out that this choice does not give the right
expression for the masses of the spinor and singlets. This means we would be
specifying a one-parameter family with the right gauge enhancements
but not the dual family of the BPS states we identified on Type
IIB. The masses of BPS states on both sides must match as well as the
gauge enhancement pattern.

To obtain the correct BPS masses in the Heterotic theory we impose the condition that it matches
the ones on Type IIB. This will fix $R$ and $B$.

The analysis naturally separate in two parts. The real part of the mass
formula depends on $B$ and $a$ only. The constraint described above
will fix $B$ in terms of $a$. The imaginary part depends on the
radius, $R$, and $a$ only. The match with the mass formulas on type
IIB fixes the relation between the two moduli. The imaginary part is
given by
\be
Im(M_h) = m_9 - A_9\cdot Q - w^9(\f{R^2}{2} + (1+a^2)).
\ee
where we used
\bea
Im(T) &=&\f{\sqrt{g}}{2} + \f{A_9^2}{2} \\
&=& \f{R^2}{2} + (1+a^2) .
\eea

For the real part we obtain 
\be
Re(M_h) = m_8 - A_8\cdot Q +w^9(\f{B}{2} -\1/2) .
\ee
where
\bea
Im(T) &=&\f{B}{2} - \f{A_8\cdot A_9}{2} \\
&=&\f{B}{2} - \1/2  .
\eea

Now we require that the imaginary part of the masses for both the spinors and
singlets to be equal to $(1/2 - a)$ in order to agree with the map for the
vectors as in Table 6. We choose three specific BPS gauge bosons to
carry out the analysis explicitly. The results can be easily extended to
all other gauge fields. The representatives we choose are
\bea
Q^1 &=& (0^4,-1^4,0^2,-1^6) \in ({\bf 1,1,1,1,1}) \\
Q^2 &=& (+\1/2^4,-\1/2^4,+\1/2^2,-\1/2^6) \in ({\bf 8_s,1,1,1,1}) \\
Q^3 &=& (+\1/2,-\1/2^3,-\1/2^4,+\1/2, -\1/2,-\1/2^6) \in ({\bf 8_c,1,2,1,1}) .
\eea
 
First of all we see from eq(B.5) that for this gauge bosons to be massless at
$a=1/2$ we must have $m_9=-1$ for the spinors and $m_9=-2$ for the
singlet. Furthermore, requiring agreement with the imaginary part of the
masses on the Type IIB theory we obtain 
\be
\f{R^2}{2} = 2a -a^2 -\f{1}{2} = \f{R_c^2}{2} - (\f{1}{2} -a) .
\ee
This fixes the radius as a function of the Wilson line parameter. 

The requirement that the real part is as in Table 6 yields for $B$
\bea
Re(M_h)(Q^1)&=&m_8+B\biggl|_{a=\1/2}=0 \\
Re(M_h)(Q^2)&=&(m_8-1+\f{B}{2})\biggl|_{a=\1/2}=0 \\
Re(M_h)(Q^3)&=&(m_8+1+\f{B}{2})\biggl|_{a=\1/2}=0 .
\eea
These equations imply that $B\in 2Z$. We choose $B=0$ (up to $SL(2,Z)$
transformations). The $m_8$ quantum number for the representatives
as well as the expression for the real part of the masses away for
$a=1/2$ are also fixed. We summarise the results in table 8.

\begin{table}[hth]
\begin{center}
\begin{tabular}{|l|l|l|l|l|l|l|} \hline 
$Q$ & $m_8$ & $m_9$&$w^8$&$w^9$&$Re(M_h)$&$Im(M_h)$   \\ \hline
$Q^1$ & $0$ & $-2$&$0$&$2$&$0$&$(\1/2 -a)$ \\ 
$Q^2$ & $1$ & $-1$&$0$&$1$&$-(\1/2-a)$&$(\1/2 -a)$ \\
$Q^3$ & $-1$ & $-1$&$0$&$1$&$0$&$(\1/2 -a)$  
\\ \hline 
\end{tabular}
\end{center}
\caption{The quantum numbers and masses of the singlet and spinor representatives.}
\end{table}

\pagebreak

\end{document}